\documentclass[journal]{IEEEtran}

\usepackage{latexsym}
\usepackage{graphicx}
\usepackage{amsfonts,amssymb,amsmath}
\usepackage{ctable} 
\usepackage{mathtools, cuted}
\usepackage{hyperref}
\usepackage[ruled,linesnumbered,noend]{algorithm2e}
\usepackage[T1]{fontenc}
\usepackage{cite}
\usepackage{subcaption}
\usepackage{comment}
\usepackage{diagbox}

\usepackage{amsthm}
\usepackage{overpic}
\usepackage{steinmetz}
\usepackage{array}
\usepackage{url}
\usepackage{color}

\usepackage{algorithmicx}
\usepackage{algcompatible}

\usepackage{xcolor}
\usepackage{soul}

\usepackage{linegoal}

\theoremstyle{plain}

\newtheorem{remark}{Remark}

\newcommand{\vect}[1]{\mathbf{#1}}
\newcommand{\maximize}[1]{{\underset{{#1}}{\mathrm{maximize}}}}
\newcommand{\minimize}[1]{{\underset{{#1}}{\mathrm{minimize}}}}

\newcommand*{\LongState}[1]{\STATE
\parbox[t]{0.9\linewidth-\algorithmicindent-\algorithmicindent}{#1\strut}}

\def\Htran{\mbox{\tiny $\mathrm{H}$}}
\def\Ttran{\mbox{\tiny $\mathrm{T}$}}
\def\CN{\mathcal{C}\mathcal{N}} 

\SetCommentSty{mycommfont}

\IEEEoverridecommandlockouts

\begin{document}
\bstctlcite{BSTcontrol}
\title{Joint and Streamwise Distributed MIMO Satellite Communications with Multi-Antenna Ground Users}

\author{\IEEEauthorblockN{\normalsize Parisa Ramezani, \textit{Member, IEEE} and Emil Bj\"{o}rnson, \textit{Fellow, IEEE}, }
\thanks{
Part of this paper was presented at 2026 IEEE International Conference on Communications (ICC) Workshops \cite{Ramezani2026Cell-Free}. The authors are with the Department of Communication Systems, KTH Royal Institute of Technology, 100 44 Stockholm, Sweden (email:\{parram,emilbjo\}@kth.se). This work is funded by the SSF research center SMART-6GSAT (Sustainable Mobile Autonomous and Resilient 6G SatCom), reg.nr. CSG23-0001. }}

\maketitle
\begin{abstract}
We consider low Earth orbit downlink communication, in which multiple satellites serve multi-antenna ground users, each transmitting multiple spatial streams per user. Using a line-of-sight-dominant satellite channel model with statistical channel state information (sCSI), including angular information and large-scale fading, we study two distributed transmission modes with different fronthaul requirements. First, for joint transmission, where all satellites transmit all user streams, we formulate a sum spectral efficiency (SE) maximization problem under general convex power constraints and address the intractability of the exact ergodic SE expression by adopting a tractable surrogate SE expression. Exploiting the equivalence between sum SE maximization and weighted sum mean square error minimization, we derive a novel iterative transceiver design. Second, to reduce fronthaul load, we propose streamwise transmission, in which each stream is sent by a single satellite, and we develop an eigenmode-based stream-satellite association using participation factors and a maximum-weight bipartite matching problem solved by the Hungarian algorithm. Numerical simulations quantify the fronthaul savings of streamwise transmission, and reveal how channel orthogonality, stream/user loading, and hybrid joint-streamwise transmission affect the performance-overhead trade-off. The results also demonstrate fast convergence, clear gains over benchmark precoding and association methods, and robustness against sCSI aging and angular estimation errors.
\end{abstract}


\section{Introduction}
The realization of ubiquitous connectivity is formally identified by the International Telecommunication Union (ITU) as a central objective for future 6G networks, with the overarching goal of closing the global digital divide. Realizing this vision requires new solutions that can deliver reliable and seamless broadband access in remote and sparsely populated regions, where deploying and maintaining terrestrial infrastructure is often economically or geographically impractical. Within this framework, satellite communication systems are expected to serve a foundational role in constructing a truly global and inclusive 6G ecosystem, complementing ground deployments by extending coverage beyond the reach of conventional networks \cite{perez2019signal,Jamshed2025Non}. In particular, low Earth orbit (LEO) satellite systems have gained substantial momentum thanks to several advantages over geostationary (GEO) and medium Earth orbit (MEO) satellites, including significantly lower latency, propagation losses, and production and launch costs \cite{Wu2024Space}. 

Satellite links are inherently power-limited due to the long propagation distance, so high-gain antennas and beamforming have always been essential to concentrate radiated energy. Traditionally, beamforming has often relied on fixed analog beam patterns.
Recently, the application of multiple-input multiple-output (MIMO) techniques to satellite communications has attracted growing interest, as it enables more efficient use of spatial resources and can significantly improve link performance. By exploiting multiple antennas for transmit and/or receive beamforming, satellite MIMO can concentrate energy toward intended users, mitigate interference, and improve coverage.
Reference~\cite{you2020massive} studies massive MIMO LEO satellite downlink/uplink transmission with full frequency reuse and proposes low-complexity precoding/combining designs based on statistical CSI (sCSI) to handle inter-beam interference without relying on instantaneous CSI. Building on this direction, \cite{you2022hybrid} develops an energy-efficient hybrid precoding design for massive MIMO LEO downlink transmission to reduce the hardware cost and power consumption of fully digital arrays.
The authors of~\cite{alsenwi2024robust} propose a robust hybrid precoding scheme for massive MIMO LEO downlink transmission under highly dynamic and outdated CSI, using a risk-aware deep reinforcement learning framework to maximize energy efficiency while satisfying per-user rate constraints.
Reference~\cite{Li2024Ergodic} studies massive MIMO LEO uplink transmission with multiple multi-antenna users and derives a lower-dimensional equivalent formulation for optimal transmit covariance design. In~\cite{Darya2025Semi}, a decision-directed semi-blind channel estimation method is proposed to mitigate Doppler-induced channel aging by refining pilot-based estimates using detected data symbols.

Taking this a step further, MIMO can be realized over a distributed aperture by treating multiple satellites as a virtual antenna array. The concept of cell-free massive MIMO, long studied in terrestrial networks \cite{Ngo2017Cell,demir2021foundations,Mohammadi2023Network}, is now gaining traction in satellite communications for serving ground users through such distributed transmission.
By coordinating transmission across multiple satellites, this approach can improve coverage, increase system capacity, and provide a more uniform user experience. However, the transmission schemes must be designed differently from terrestrial systems because one can hardly achieve phase-coherent joint transmission from multiple satellite transmitters, as the differences in propagation delays and phase jitter are too large.
Hence, in this paper, we will develop novel non-coherent transmission modes tailored for satellite scenarios.

\subsection{Related Work}

Cell-free massive MIMO was introduced as a distributed terrestrial architecture in which many access points jointly serve users over the same time-frequency resources, thereby improving coverage and service uniformity compared with collocated massive MIMO~\cite{Ngo2017Cell}. Inspired by this concept, cooperative multi-satellite transmission is being explored to exploit satellite diversity and improve the performance of ground users.

Reference~\cite{Richter2020Downlink} studies two-satellite cooperative downlink transmission to a multi-antenna ground user and analyzes the effective channel and outage probability under random user orientation. In \cite{humadi2024distributed}, a user-centric distributed massive MIMO architecture for LEO constellations is proposed, where each user is served by a dynamically selected satellite cluster, with reference-satellite coordination and phase-aware precoding used to address scalability and asynchronous reception.
 In the coordinated multi-satellite multi-beam downlink considered in~\cite{ha2024user}, LEO satellites serve single-antenna users through predefined DFT-based beams, with user-satellite association, beam activation, and precoding jointly optimized to minimize transmit power subject to signal-to-interference-plus-noise ratio (SINR) constraints. Distributed beamforming for direct LEO satellite-to-smartphone links is studied in~\cite{xu2024enhancement}, where an electromagnetic-wave-based model is used to characterize the dependence of the combining gain on the relative satellite geometry.
 A dual-function LEO constellation for joint communication and sensing is proposed in~\cite{wang2025multiple}, with multiple satellites cooperatively serving users and collecting target-reflected signals under per-satellite power limits. In~\cite{Zhang2025Enabling}, an sCSI-based hybrid beamforming architecture is developed for orthogonal frequency division multiplexing (OFDM)-based cooperative LEO downlink transmission with inter-satellite links.

These studies highlight the promising potential of joint transmission from multiple satellites to ground users, demonstrating that cooperation can strengthen weak links and improve performance through joint beamforming and interference management.  
However, they focus on single-antenna ground users, which inherently restricts the number of spatial degrees of freedom available at the receiver and therefore limits spatial multiplexing and stream-level processing.
As state-of-the-art terrestrial and satellite receivers are typically equipped with multiple antennas, it is important to study satellite communication systems with multi-antenna ground users and exploit the additional spatial degrees of freedom to support multi-stream transmission and spatial interference cancellation.

\subsection{Contributions}
This paper investigates a distributed downlink satellite communication system, where multiple satellites serve multi-antenna ground users by transmitting multiple spatial data streams to each user. Unlike conventional cell-free networks, where distributed access points can be tightly synchronized to coherently transmit the same data symbols to the users, such coherent joint transmission is difficult to realize in distributed LEO satellite systems due to the large inter-satellite separation, different propagation delays, Doppler shifts, and rapidly varying relative phases. In light of this, we consider two transmission modes: joint non-coherent transmission, where all satellites transmit all streams to each user in a non-coherent manner, and streamwise transmission, where each stream is assigned to a single satellite to reduce the fronthaul load.

 The contributions of this paper can be summarized as follows:
\begin{itemize}
    \item We formulate the joint non-coherent precoder design for distributed multi-satellite MIMO downlink as a sum SE maximization problem under general convex power constraints. To avoid the intractable exact ergodic SE, we adopt an sCSI-based surrogate that is independent of instantaneous channel phases, enabling precoder design without inter-satellite phase synchronization. Using the weighted minimum mean square error (WMMSE) equivalence~\cite{Shi2011}, we derive a block coordinate descent algorithm that updates the receive combiners, MSE weights, and transmit precoders, with the Lagrange multipliers of the power constraints obtained via the ellipsoid method.

   \item We propose a fronthaul-efficient streamwise transmission mode, where each data stream is transmitted by only one satellite. We formulate the stream-satellite association based on eigenmode participation factors and solve the resulting maximum-weight bipartite matching problem using the Hungarian algorithm.

   \item Through extensive numerical simulations, we characterize the validity of the surrogate SE expression, the convergence of the proposed algorithms, and the performance-fronthaul trade-off between joint and streamwise transmission.  We further quantify the fronthaul savings, study the impact of user/stream loading, examine a hybrid transmission strategy, and evaluate the robustness of the proposed schemes against sCSI aging and angular estimation errors.

\end{itemize} 

 This work is an extended and refined version of \cite{Ramezani2026Cell-Free}. Compared to the conference paper, which focuses on joint multi-satellite transmission and considers per-satellite and per-antenna power constraints, this paper develops a unified precoding framework under general convex power constraints, thereby accommodating a wide range of practically relevant constraints. 
A key additional contribution is the proposed streamwise mode, together with an eigenmode-based stream-satellite pairing algorithm, that reduces fronthaul load while still taking advantage of multi-satellite multi-stream transmission. Overall, we broaden the problem formulation and analysis, introduce new algorithms, and expand the numerical evaluations to provide deeper insights into the performance-overhead trade-offs of distributed multi-satellite transmission with multi-antenna ground users receiving multiple spatial streams.

\subsection{Organization}
 The rest of this paper is structured as follows. Section~\ref{sec:sysmod} presents the system model, formulates the sum SE maximization problem, and introduces an SE approximation that makes the problem tractable.  Section~\ref{sec:joint_precoding} details the proposed precoding design for non-coherent joint transmission where all satellites transmit multiple streams to all the ground users. In Section~\ref{sec:streamwise}, we propose a lower-overhead distributed transmission scheme where each user stream is served by a single satellite, and develop a stream-satellite association strategy. The computational complexity analysis of the proposed schemes is provided in Section~\ref{sec:complexity}.   Section~\ref{sec:results} provides supporting numerical simulations, and Section~\ref{sec:conclusions} concludes the paper.

\subsection{Notations} Scalars are denoted by italic letters, while vectors and matrices are denoted by bold-face lower-case and upper-case letters, respectively. $(\cdot)^{\Ttran}$ and $(\cdot)^{\Htran}$ indicate the transpose and conjugate transpose, respectively, and $\Re(\cdot)$ denotes the operation of taking the real part. For the vector $\vect{x}$, $\|\vect{x}\|$ is its Euclidean norm and $\vect{x} \succ \vect{0}$ means that all entries of $\vect{x}$ are strictly positive. 
For the matrix $\vect{X}$, $\|\vect{X}\|_F$ represents its Frobenius norm and $\vect{X} \succeq \vect{0}$ implies that $\vect{X}$ is positive semidefinite.  $|\vect{X}|$ and $\mathrm{Tr}(\vect{X})$ respectively return the determinant and trace of matrix $\vect{X}$. $\mathbb{C}$ denotes the set of complex numbers.  $\CN(0,\omega)$ represents a circularly symmetric complex Gaussian distribution with mean $0$ and variance $\omega$, and $\CN(\vect{0}, \boldsymbol{\Omega})$ denotes its multivariate counterpart with mean vector $\vect{0}$ and covariance matrix $\boldsymbol{\Omega}$. 
The operator $\odot$ is the elementwise product, and $\vect{X}(i,i)$ indicates the $i$-th diagonal entry of $\vect{X}$. $[x]_+$ denotes clipping at zero, i.e., $[x]_+ = \max\{0,x\}$. 
\begin{figure}
    \centering
    \includegraphics[width=0.85\linewidth]{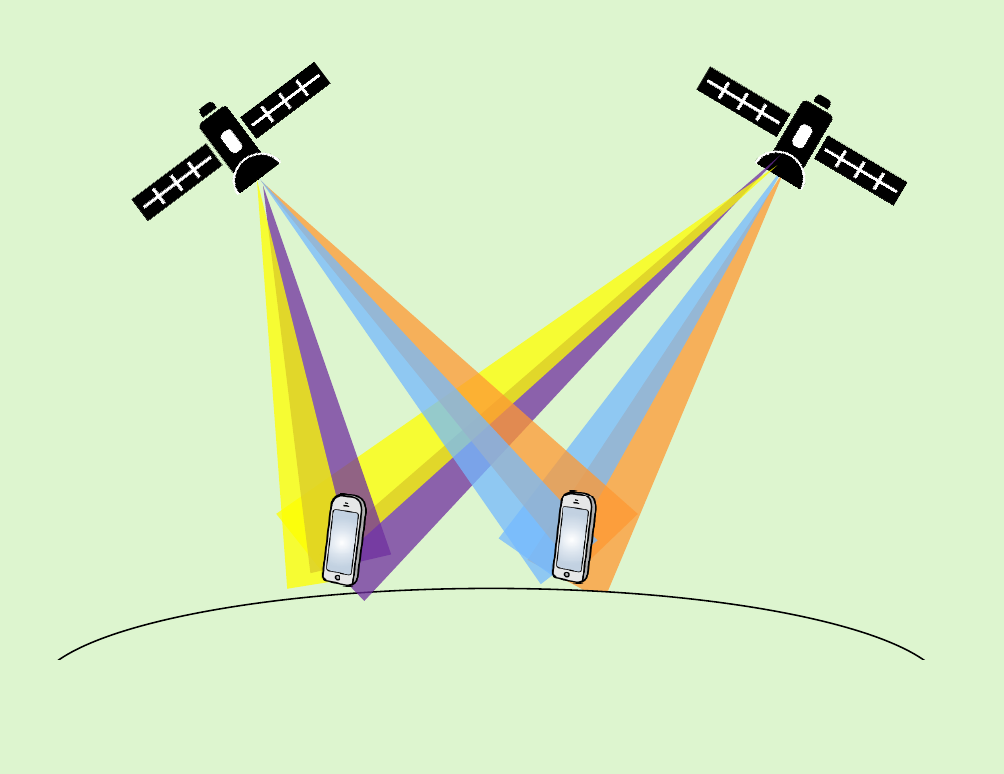}
    \caption{Distributed multi-satellite MIMO downlink with multi-stream transmission to multi-antenna users. }
\label{fig:satcom}
\end{figure}
\section{System Model and Problem Formulation}
\label{sec:sysmod}
As shown in Fig. ~\ref{fig:satcom}, a downlink multi-user distributed satellite communication system is considered where a constellation of $L$ LEO satellites (SATs), each equipped with $N$ antennas, serve $K$ user equipments (UEs), each with $M$ antennas. Since the UEs have multiple antennas, multiple spatially multiplexed data streams can be sent to each UE, if the joint channel from the SATs supports it. Let $\vect{x}_k \sim \CN (\vect{0},\vect{I}_S) \in \mathbb{C}^{S}$ denote the vector of $S \leq M$ data streams for UE~$k$. The signals for different UEs are independent
from each other and from the receiver noises. With joint transmission, the received signal at UE~$k$ is
\begin{align}
\label{eq:received_signal_1}
  \vect{y}_k = \sum_{l=1}^L \vect{H}_{l,k} \vect{W}_{l,k} \vect{x}_k + \sum_{i = 1, i \neq k}^K \sum_{l=1}^L \vect{H}_{l,k}\vect{W}_{l,i}\vect{x}_i + \vect{n}_k,   
\end{align}
where $\vect{H}_{l,k} \in \mathbb{C}^{M \times N}$ is the channel between SAT~$l$ and UE~$k$, $\vect{W}_{l,k} \in \mathbb{C}^{N \times S }$ is the precoding matrix used by SAT~$l$ for UE~$k$, and $\vect{n}_k \sim \CN(0, \sigma^2\vect{I}_M)$ is the noise vector at UE~$k$.

\subsection{Channel Model}
We consider the typical scenario in satellite communications: far-field line-of-sight (LoS) channels between the SATs and UEs. We further assume that the antenna panels at the SATs and UEs are in the form of uniform linear arrays (ULAs). Under these assumptions, the channel $\vect{H}_{l,k}$ can be modeled as 
\begin{align}
\label{eq:geometric_channel}
  \vect{H}_{l,k} = \gamma_{l,k} \vect{b}_k(\theta_{l,k})\vect{a}_l^{\Ttran}(\varphi_{l,k}),  
\end{align}
where $\gamma_{l,k}$ is the complex channel gain between SAT~$l$ and UE~$k$, $\vect{b}_k(\cdot)$ and $\vect{a}_l(\cdot)$ are respectively the array response vectors at UE~$k$ and SAT~$l$, and $\theta_{l,k}$ and $\varphi_{l,k}$ are the respective angle-of-arrival (AoA) and angle-of-departure (AoD).  The SATs typically follow precise deterministic orbital trajectories and the UEs' locations can be reliably tracked via GPS. Thus, it is reasonable to assume that the AoA and AoD information is available at the SATs and UEs \cite{you2020massive, you2022hybrid, Li2024Ergodic, Wang2025Statistical, Zhang2025Enabling}. However, the channel gain $\gamma_{l,k}$
 varies rapidly; thus, only its statistical information is usually available at the transmitter. 
 We model $\gamma_{l,k}$ as Rician fading, where the LoS component has a time-varying random phase to capture rapid phase fluctuations, and the non-LoS (NLoS) component represents scattering around the UE. Specifically, 
 \begin{equation}
  \gamma_{l,k} = \sqrt{\beta_{l,k}} \left(\sqrt{\frac{\kappa_{l,k}}{\kappa_{l,k}+1}} e^{j\psi_{l,k}} + \sqrt{\frac{1}{\kappa_{l,k} + 1}} z_{l,k} \right),   
 \end{equation}
 where $\psi_{l,k} \sim \mathrm{Uniform}[0,2\pi)$, $z_{l,k} \sim \CN(0,1)$, $\psi_{l,k}$ and $z_{l,k}$ are independent for all $l,k$, and $\beta_{l,k}$ is known a priori.

\subsection{Sum SE Maximization Problem Formulation}

 Since $\gamma_{l,k}$ varies rapidly over time and frequency, we adopt the ergodic achievable SE as the performance metric, assuming coding over many fading realizations. 
Based on the received signal expression in \eqref{eq:received_signal_1}, the ergodic achievable SE at UE~$k$ is given by \eqref{eq:achievable_rate}, shown at the top of the next page, where the expectation is over different channel realizations.   
\begin{figure*}[t]
    \begin{align}
    \label{eq:achievable_rate}
  R_k = \mathbb{E}\left\{\log_2 \left| \vect{I}_M + \left(\sum_{l=1}^L \vect{H}_{l,k} \vect{W}_{l,k}\right) \left(\sum_{l=1}^L \vect{H}_{l,k} \vect{W}_{l,k}\right)^{\Htran} \left(\sum_{i=1,i\neq k}^K \left(\sum_{l=1}^L \vect{H}_{l,k}\vect{W}_{l,i} \right)\left(\sum_{l=1}^L \vect{H}_{l,k}\vect{W}_{l,i} \right)^{\Htran}   + \sigma^2\vect{I}_M\right)^{-1}\right| \right\}.  
\end{align}
    \hrulefill
\end{figure*}
The objective in this paper is to maximize the sum SE of all UEs by optimizing the precoding matrices at each SAT, subject to a number of convex power constraints. 
The problem of interest is thus formulated as 
\begin{subequations}
\label{eq:main_problem}
 \begin{align}
  &\maximize{\{\vect{W}_{l,k}\}_{\forall l,k}}\,\, \sum_{k=1}^K R_k, \\
    &\mathrm{subject~to}\, \sum_{k=1}^K \mathrm{Tr}\left(\vect{W}_{l,k}^{\Htran} \vect{A}_{l,x}\vect{W}_{l,k} \right) \leq \rho_{l,x},\,\forall l, x = 1,\ldots,X_l, \label{eq:power_constraint}
    \end{align}
 \end{subequations}
where $\vect{A}_{l,x} \in \mathbb{C}^{N \times N}\succeq \vect{0}$ is a given weighting matrix that defines the $x$-th convex power constraint for SAT~$l$ and limits the power in a particular signal subspace of SAT~$l$, and $\rho_{l,x}$ is the corresponding power cap for that subspace. 
The power constraints in \eqref{eq:power_constraint} represent a general model that includes many standard power constraints used in the literature. For example, the per-SAT total power constraint is obtained by setting only one constraint for each SAT ($X_l = 1~\forall l$) with $\vect{A}_{l,1} = \vect{I}_N$. Also, per-antenna power constraints are obtained by having $X_l = N~\forall l$, and choosing $\vect{A}_{l,n} = \vect{D}_n$ where $\vect{D}_n$ is a diagonal matrix with $1$ on its $n$-th diagonal element and $0$ elsewhere. 
$R_k$ does not have an analytical form, which makes problem \eqref{eq:main_problem} difficult to handle. 
In the following, we introduce a tractable surrogate function for the achievable SE in \eqref{eq:achievable_rate}, which will be used to reformulate \eqref{eq:main_problem}.

\subsection{Surrogate Function}
\label{sec:surrogate}
As the first step to solve \eqref{eq:main_problem}, we seek to reformulate the objective function in a more tractable form.  For positive definite matrices $\vect{X}$ and $\vect{Y}$, we use the surrogate 
\begin{align}
\label{eq:approximation}
     \Phi(\vect{X},\vect{Y}) \triangleq \log_2 \left|\vect{I} + \mathbb{E}\{\vect{X}\} \mathbb{E}\{\vect{Y}\}^{-1}  \right|,
  \end{align}
  for the intractable ergodic expression $\mathbb{E}\left\{\log_2\left|\vect{I}+\vect{XY}^{-1}\right|\right\}$. This surrogate allows us to replace the random desired signal and interference-plus-noise covariance matrices by their statistical averages. The expression in \eqref{eq:approximation} has been used in some earlier works (e.g., \cite{zhou2024robust}) to approximate the ergodic rate expression. While the approximation has been shown to be tight in scalar cases \cite{zhang2014power}, to the best of our knowledge, its tightness cannot be established under general matrix-valued settings. We next discuss some relevant special cases that provide further insight into the behavior of this surrogate.

  Starting from Jensen's inequality, we have
  \begin{equation}
  \label{eq:Jensen}
    \mathbb{E}\left\{\log_2\left|\vect{I}+\vect{XY}^{-1}\right|\right\} \leq \log_2 \left|\vect{I} + \mathbb{E}\{\vect{X}\vect{Y}^{-1}\}\right|.  
  \end{equation}
  When it comes to achievable SE, we have $\mathbb{E}\{\vect{X} \vect{Y}^{-1}\} \approx \mathbb{E}\{\vect{X}\} \mathbb{E}\{\vect{Y}\}^{-1}$
  in regimes where the interference-plus-noise covariance $\vect{Y}$ exhibits limited randomness. This occurs, for example, when noise dominates the interference term, or when the designed precoders effectively suppress interference leakage toward unintended UEs so that the residual interference covariance is small and fluctuates weakly around its mean. In such cases, \eqref{eq:Jensen} leads to 
  \begin{equation}
  \label{eq:upper-bound}
   \mathbb{E}\left\{\log_2\left|\vect{I}+\vect{XY}^{-1}\right|\right\} \leq \log_2 \left|\vect{I} + \mathbb{E}\{\vect{X}\} \mathbb{E}\{ \vect{Y}\}^{-1}\right|,   
  \end{equation}
and the surrogate function in \eqref{eq:approximation} can be interpreted as a tractable upper bound on the ergodic achievable SE. 

  Moreover, the surrogate becomes tight with respect to the exact ergodic expression in the low transmit power regime. In particular, at low transmit power, the interference term in $\vect{Y}$ is negligible compared with the noise covariance. Hence, $\vect{Y}\approx \sigma^2\vect{I}_M$, and we have
\begin{align}
\mathbb{E}\left\{\log_2\left|\vect{I}+\vect{X}\vect{Y}^{-1}\right|\right\}
&\approx
\mathbb{E}\left\{\log_2\left|\vect{I}+\frac{1}{\sigma^2}\vect{X}\right|\right\}. \label{eq:low_power_step_1}
\end{align}
Furthermore, since the desired signal covariance $\vect{X}$ is also small at low transmit power, we can use the first-order expansion $\log_2|\vect{I}_M+\vect{A}|
\approx
\frac{1}{\ln 2}\mathrm{Tr}(\vect{A})$ which yields
\begin{align}
\mathbb{E}\left\{\log_2\left|\vect{I}_M+\frac{1}{\sigma^2}\vect{X}\right|\right\}
&\approx
\mathbb{E}\left\{
\frac{1}{\ln 2}
\mathrm{Tr}\left(\frac{1}{\sigma^2}\vect{X}\right)
\right\} \notag\\
&=
\frac{1}{\ln 2}
\mathrm{Tr}\left(\frac{1}{\sigma^2}\mathbb{E}\{\vect{X}\}\right). \label{eq:low_power_step_2}
\end{align}
In the same low transmit power regime, we also have
$\mathbb{E}\{\vect{Y}_k\}\approx \sigma^2\vect{I}_M$, resulting in $\mathbb{E}\{\vect{Y}_k\}^{-1}\approx \frac{1}{\sigma^2}\vect{I}_M.$
Therefore,
\begin{align}
\frac{1}{\ln 2}
\mathrm{Tr}\left(\frac{1}{\sigma^2}\mathbb{E}\{\vect{X}\}\right)
&\approx
\frac{1}{\ln 2}
\mathrm{Tr}\left(\mathbb{E}\{\vect{X}\}\mathbb{E}\{\vect{Y}\}^{-1}\right). \label{eq:low_power_step_3}
\end{align}
Finally, since \(\mathbb{E}\{\vect{X}\}\mathbb{E}\{\vect{Y}\}^{-1}\) is also small at low transmit power, applying $\log_2|\vect{I}_M+\vect{A}|
\approx
\frac{1}{\ln 2}\mathrm{Tr}(\vect{A})$ in reverse gives
\begin{align}
\frac{1}{\ln 2}
\mathrm{Tr}\left(\mathbb{E}\{\vect{X}\}\mathbb{E}\{\vect{Y}\}^{-1}\right)
&\approx
\log_2\left|
\vect{I}+\mathbb{E}\{\vect{X}\}\mathbb{E}\{\vect{Y}\}^{-1}
\right|. \label{eq:low_power_step_4}
\end{align}
Therefore, at low transmit power, the following approximation holds:
\begin{equation}
\label{eq:approximation_low_power}
\mathbb{E}\left\{\log_2\left|\vect{I}+\vect{X}\vect{Y}^{-1}\right|\right\}
\approx
\log_2\left|
\vect{I}+\mathbb{E}\{\vect{X}\}\mathbb{E}\{\vect{Y}\}^{-1}
\right|.
\end{equation} 
In Section~\ref{sec:results}, we will numerically show that the surrogate function behaves as an upper bound for the exact ergodic expression under the considered regular operating conditions, while it becomes a tight approximation of the exact expression in the low transmit power regime.
 \subsection{Problem Reformulation} 
 The SE expression can be simplified using \eqref{eq:approximation}. In particular, for the signal term, we have 
 \begin{align}
 \label{eq:signal_approx}
  &\mathbb{E}\left\{\left(\sum_{l=1}^L \vect{H}_{l,k} \vect{W}_{l,k}\right) \left(\sum_{l=1}^L \vect{H}_{l,k} \vect{W}_{l,k}\right)^{\Htran}\right\}  \nonumber \\
  &  = \mathbb{E}\left\{\sum_{l=1}^L |\gamma_{l,k}|^2 \vect{b}_{l,k} \vect{a}_{l,k}^{\Ttran} \vect{W}_{l,k} \vect{W}_{l,k}^{\Htran} \vect{a}_{l,k}^* \vect{b}_{l,k}^{\Htran}\right\} \nonumber \\
  &  = \sum_{l=1}^L \mathbb{E}\left\{|\gamma_{l,k}|^2\right\} \vect{b}_{l,k}\vect{a}_{l,k}^{\Ttran} \vect{W}_{l,k} \vect{W}_{l,k}^{\Htran} \vect{a}_{l,k}^* \vect{b}_{l,k}^{\Htran} \nonumber \\
  & = \sum_{l=1}^L \beta_{l,k} \vect{b}_{l,k}\vect{a}_{l,k}^{\Ttran} \vect{W}_{l,k} \vect{W}_{l,k}^{\Htran} \vect{a}_{l,k}^* \vect{b}_{l,k}^{\Htran},
 \end{align}
where we use the notations $\vect{b}_{l,k} = \vect{b}_k(\theta_{l,k})$ and $\vect{a}_{l,k} = \vect{a}_k(\varphi_{l,k})$. For the interference-plus-noise term, we obtain
\begin{align}
\label{eq:interference_approx}
 &\mathbb{E}\left\{\sum_{i=1,i\neq k}^K \left(\sum_{l=1}^L \vect{H}_{l,k}\vect{W}_{l,i} \right)\left(\sum_{l=1}^L \vect{H}_{l,k}\vect{W}_{l,i} \right)^{\Htran}   + \sigma^2\vect{I}_M  \right\}  \nonumber \\ 
 & = \sum_{i=1,i\neq k}^K \sum_{l=1}^L \beta_{l,k}\vect{b}_{l,k}\vect{a}_{l,k}^{\Ttran} \vect{W}_{l,i} \vect{W}_{l,i}^{\Htran} \vect{a}_{l,k}^* \vect{b}_{l,k}^{\Htran} + \sigma^2 \vect{I}_M.
\end{align}
Therefore, the SE expression $R_k$ in \eqref{eq:achievable_rate} can be replaced by a surrogate SE expression  $\bar{R}_k$ in \eqref{eq:achievable_rate_approx}, shown at the top of the next page.%
\begin{figure*}[!t]
    \begin{align}
    \label{eq:achievable_rate_approx}
  \bar{R}_k = \log_2 \left| \vect{I}_M + \left( \sum_{l=1}^L \beta_{l,k} \vect{b}_{l,k}\vect{a}_{l,k}^{\Ttran} \vect{W}_{l,k} \vect{W}_{l,k}^{\Htran} \vect{a}_{l,k}^* \vect{b}_{l,k}^{\Htran} \right) \left( \sum_{i=1,i\neq k}^K \sum_{l=1}^L \beta_{l,k}\vect{b}_{l,k}\vect{a}_{l,k}^{\Ttran} \vect{W}_{l,i} \vect{W}_{l,i}^{\Htran} \vect{a}_{l,k}^* \vect{b}_{l,k}^{\Htran} + \sigma^2 \vect{I}_M \right)^{-1} \right|. 
\end{align}
    \hrulefill
\end{figure*}

The sum SE maximization problem is thus reformulated as 
\vspace{-2mm}
\begin{subequations}
\label{eq:approximate_problem}
 \begin{align}
  &\maximize{\{\vect{W}_{l,k}\}_{\forall l,k}}\,\, \sum_{k=1}^K \bar{R}_k, \\
    &\mathrm{subject~to}\, \sum_{k=1}^K \mathrm{Tr}\left(\vect{W}_{l,k}^{\Htran}\vect{A}_{l,x}\vect{W}_{l,k} \right) \leq \rho_{l,x},~\forall l,\,x = 1,\ldots,X_l. \label{eq:power_constraints} 
    \end{align}
 \end{subequations}

 In what follows, we first focus on the joint transmission mode considered in problem~\eqref{eq:approximate_problem}, which assumes full message sharing across SATs. We then extend the proposed framework to a mode with lower fronthaul overhead that reduces the message-sharing requirement by restricting each stream to be transmitted by a single SAT through a stream-SAT association.
 
\section{Precoding Design for Non-Coherent Joint Transmission}
\label{sec:joint_precoding}  

 Problem \eqref{eq:approximate_problem} is non-convex due to the non-concavity of the objective function and the coupling between the precoding matrices. 
 The WMMSE framework is a well-established approach for solving non-convex sum SE maximization problems involving interference terms \cite{Shi2011}. In this section, we follow the WMMSE approach and tailor it to problem \eqref{eq:approximate_problem} by accommodating multiple general convex per-SAT power constraints. 
 Since the reformulated problem \eqref{eq:approximate_problem} is based on the surrogate SE expression in \eqref{eq:achievable_rate_approx}, which depends on the channel only through the deterministic effective matrices 
$\tilde{\vect{H}}_{l,k}=\sqrt{\beta_{l,k}}\vect{b}_{l,k}\vect{a}_{l,k}^{\Ttran} \in \mathbb{C}^{M \times N}$, 
we henceforth perform the derivations using the corresponding effective received signal obtained from \eqref{eq:received_signal_1} by replacing $\vect{H}_{l,k}$ with $\tilde{\vect{H}}_{l,k}$. 
This yields
\begin{equation}
\label{eq:effective_received_signal}
\Tilde{\vect{y}}_k = \sum_{l=1}^L \Tilde{\vect{H}}_{l,k} \vect{W}_{l,k} \vect{x}_k + \sum_{i = 1, i \neq k}^K \sum_{l=1}^L \Tilde{\vect{H}}_{l,k}\vect{W}_{l,i}\vect{x}_i + \vect{n}_k.   
\end{equation}
 Assuming that UE~$k$ uses the receive combining matrix $\vect{U}_k \in \mathbb{C}^{M \times S}$ to decode its desired signal, the estimated signal at UE~$k$ is given by $\hat{\vect{x}}_k = \vect{U}_k^{\Htran}\Tilde{\vect{y}}_k$.   The MSE matrix for UE~$k$ can be expressed as 
 \begin{align}
 \label{eq:MSE}
 &\vect{E}_k = \mathbb{E}\left\{ (\hat{\vect{x}}_k - \vect{x}_k)(\hat{\vect{x}}_k - \vect{x}_k)^{\Htran} \right\} = \nonumber \\
 & \vect{U}_k^{\Htran} \left (\sum_{i=1}^K \sum_{l=1}^L \Tilde{\vect{H}}_{l,k} \vect{W}_{l,i} \vect{W}_{l,i}^{\Htran} \Tilde{\vect{H}}_{l,k}^{\Htran}\right) \vect{U}_k \nonumber \\ &-2\Re\left(\vect{U}_k^{\Htran} \sum_{l=1}^L \Tilde{\vect{H}}_{l,k}\vect{W}_{l,k}\right) + \sigma^2 \vect{U}_k^{\Htran} \vect{U}_k + \vect{I}_S.
 \end{align}
 
 The sum SE maximization problem in \eqref{eq:approximate_problem} can now be formulated as an equivalent weighted sum MSE minimization problem following the approach from \cite{Shi2011}. Specifically, letting $\vect{C}_k \in \mathbb{C}^{S \times S}$ be a weight matrix associated with $\vect{E}_k$, we can solve the following problem instead of \eqref{eq:approximate_problem}:  
 \begin{subequations}
\label{eq:WMMSE_problem}
 \begin{align}
  &\minimize{\{\vect{U}_k\}_{\forall k}, \{\vect{C}_k\}_{\forall k}, \{\vect{W}_{l,k}\}_{\forall l,k}}\,\, \sum_{k=1}^K \mathrm{Tr}\left(\vect{C}_k \vect{E}_k \right) - \log_2 \left|\vect{C}_k \right|, \label{eq:WMMSE_OF}\\
    &\mathrm{subject~to}\, \sum_{k=1}^K \mathrm{Tr}\left(\vect{W}_{l,k}^{\Htran}\vect{A}_{l,x}\vect{W}_{l,k} \right) \leq \rho_{l,x},\,\forall l,\,x = 1,\ldots,X_l. 
    \end{align}
 \end{subequations}
 Problem \eqref{eq:WMMSE_problem} is equivalent to problem \eqref{eq:approximate_problem} in the sense that the optimal precoding matrices $\left\{\vect{W}_{l,k}\right\}$ are the same for both problems. Yet, \eqref{eq:WMMSE_problem} is more tractable than \eqref{eq:approximate_problem} as it is convex with respect to each block of variables, i.e., $\{\vect{U}_k\}$, $\{\vect{C}_k\}$, and $\{\vect{W}_{l,k}\}$. 
 Specifically, for fixed transmit precoders and MSE weights, the terms depending on the receive combiners are quadratic, and the associated covariance matrix is positive semidefinite; hence, the receive combiner update is a convex quadratic problem. For fixed receive combiners and transmit precoders, the objective is convex with respect to the positive definite MSE weight matrices, since it consists of a linear trace term and the convex function $-\log_2|\cdot|$. Finally, for fixed receive combiners and MSE weights, the objective is a convex quadratic function of the transmit precoders, while the power constraints are also convex quadratic constraints.
 This motivates the use of the block coordinate descent method since the per-block convexity ensures convergence to a stationary point \cite{Tseng2001}.  

 Assuming fixed $\{\vect{W}_{l,k}\}$, the optimal receive combining matrix can be readily obtained by setting $\frac{\partial \mathrm{Tr}( \vect{E}_k)}{\partial \vect{U}_k} = \vect{0}$ for each individual UE. This results in
 \begin{align}
 \label{eq:receive_beamforming}
  \vect{U}^\star_k = \left(\sum_{i=1}^K \sum_{l=1}^L \Tilde{\vect{H}}_{l,k} \vect{W}_{l,i} \vect{W}_{l,i}^{\Htran} \Tilde{\vect{H}}_{l,k}^{\Htran} + \sigma^2 \vect{I}_M \right)^{-1} \Tilde{\vect{H}}_k \vect{W}_i.
 \end{align}
Then, assuming $\{\vect{U}_k\}$ and $\{\vect{W}_{l,k}\}$ are fixed, the optimal weight matrix $\vect{C}_k$ can be found by solving $\frac{\partial \left( \mathrm{Tr}\left(\vect{C}_k \vect{E}_k - \log_2 |\vect{C}_k|\right)\right)}{\partial \vect{C}_k} = \vect{0}$ for $\vect{C}_k$, resulting in
\begin{align}
\label{eq:weight_matrix}
  \vect{C}_k^\star = \frac{1}{\mathrm{ln}\,2}\, \vect{E}_k^{-1},  
\end{align}
where $\vect{E}_k$ is obtained from \eqref{eq:MSE} after substituting $\vect{U}_k^\star$.
 The final step is to optimize $\{\vect{W}_{l,k}\}$, which can be decomposed into $L$ sub-problems with the $l$-th sub-problem being
 \begin{subequations}
\label{eq:precoding_optimization_separated}
 \begin{align}
  &\minimize{\{\vect{W}_{l,k}\}_{\forall l,k}}\,\, \sum_{k=1}^K \mathrm{Tr}\Bigg(\sum_{i=1}^K 
\vect{C}_i \vect{U}_i^{\Htran}   
\Tilde{\vect{H}}_{l,i} \vect{W}_{l,k} \vect{W}_{l,k}^{\Htran} 
\Tilde{\vect{H}}_{l,i}^{\Htran} \vect{U}_i   \nonumber \\
& - 2\vect{C}_k\Re\left(\vect{U}_k^{\Htran}  
\Tilde{\vect{H}}_{l,k}\vect{W}_{l,k}\right) \Bigg), \\
    &~\mathrm{subject~to}\, ~\sum_{k=1}^K \mathrm{Tr}\left(\vect{W}_{l,k}^{\Htran} \vect{A}_{l,x}\vect{W}_{l,k} \right) \leq \rho_{l,x},\,x = 1,\ldots, X_l. \label{eq:linear_power_constraints}
    \end{align}
 \end{subequations}
 We can now utilize the method of Lagrange multipliers to solve \eqref{eq:precoding_optimization_separated}. Specifically, the Lagrangian is given by 
 \begin{align}
   &\mathcal{L}_l = \sum_{k=1}^K \mathrm{Tr}\left( \vect{W}_{l,k}^{\Htran} \left(\sum_{i=1}^K \Tilde{\vect{H}}_{l,i}^{\Htran} \vect{U}_i \vect{C}_i \vect{U}_i^{\Htran} \Tilde{\vect{H}}_{l,i}\right) \vect{W}_{l,k} \right. -\nonumber \\
   &    2 \vect{C}_k \Re \left(\vect{U}_k^{\Htran}  
\Tilde{\vect{H}}_{l,k}\vect{W}_{l,k}\right) \Bigg) \nonumber \\ 
&+ \sum_{x=1}^{X_l}\left(\mu_{l,x} \left( \sum_{k=1}^K \mathrm{Tr}\left(\vect{W}_{l,k}^{\Htran} \vect{A}_{l,x}\vect{W}_{l,k}\right) - \rho_{l,x} \right)\right),
 \end{align}
where $\mu_{l,x}$ is the Lagrange multiplier associated with the $x$-th power constraint at SAT~$l$. We need to solve the following problem 
\begin{align}
   \minimize{\{\vect{W}_{l,k}\}_{\forall k}}\, \mathcal{L}_l, 
\end{align}
which can be separately solved for the $K$ UEs by setting $\frac{\partial \mathcal{L}_l}{\partial \vect{W}_{l,k}} = \vect{0}$. The optimal $\vect{W}_{l,k}$ for given $\boldsymbol{\mu}_l = [\mu_{l,1},\ldots,\mu_{l,X_l}]^{\Ttran}$ is obtained as 
\begin{align}
\label{eq:precoding_matrix}
 &\vect{W}_{l,k}^\star (\boldsymbol{\mu}_l) = \nonumber \\ &\left(\sum_{i=1}^K \Tilde{\vect{H}}_{l,i}^{\Htran} \vect{U}_i \vect{C}_i \vect{U}_i^{\Htran}\Tilde{\vect{H}}_{l,i} + \sum_{x = 1}^{X_l}\mu_{l,x} \vect{A}_{l,x}\right)^{-1} \Tilde{\vect{H}}_{l,k}^{\Htran}\vect{U}_k \vect{C}_k.    
\end{align}
To obtain the optimal Lagrange multiplier vector $\boldsymbol{\mu}_l^\star$, we first examine $\boldsymbol{\mu}_l = \vect{0}$. If the precoder $\vect{W}_{l,k}^\star (\vect{0})$ computed from \eqref{eq:precoding_matrix} satisfies all power constraints in \eqref{eq:linear_power_constraints}, then $\boldsymbol{\mu}_l^\star = \vect{0}$. Otherwise, we update $\boldsymbol{\mu}_l$ using the ellipsoid method \cite{grotschel2012geometric}. Specifically, we first identify a feasible upper-bound vector $\bar{\boldsymbol{\mu}}_l$ by geometric expansion: starting from $\boldsymbol{\mu}_l \succ \vect{0}$, we set $\boldsymbol{\mu}_l \leftarrow \upsilon \boldsymbol{\mu}_l\,(\upsilon >1)$ and recompute \eqref{eq:precoding_matrix} until all constraints in \eqref{eq:linear_power_constraints} are met. We initialize an ellipsoid that contains the box $[\vect{0},\bar{\boldsymbol{\mu}}_l]$. 
Then, at iteration $t$, we compute $\vect{W}_{l,k}^\star(\boldsymbol{\mu}_l^{(t)})$ from \eqref{eq:precoding_matrix} and evaluate the residuals 
\begin{align}
\label{eq:residuals}
  &g_{l,x}(\boldsymbol{\mu}_l^{(t)}) = \nonumber \\  &\sum_{k=1}^K \mathrm{Tr}\left(\vect{W}_{l,k} ^{\Htran}(\boldsymbol{\mu}_l^{(t)})\vect{A}_{l,x}\vect{W}_{l,k}(\boldsymbol{\mu}_l^{(t)}) \right) - \rho_{l,x},\,x = 1,\ldots,X_l,
\end{align}
where $\boldsymbol{\mu}_l^{(t)}$ is the center of the current ellipsoid. We then generate a cutting plane through the ellipsoid center using the residual vector $\vect{g}_l^{(t)}=[g_{l,1}(\boldsymbol{\mu}_l^{(t)}),\ldots,g_{l,X_l}(\boldsymbol{\mu}_l^{(t)})]^{\Ttran}$ as the cut direction. Next, we apply a central-cut ellipsoid update: the ellipsoid is shrunk, and its center is shifted opposite to the normalized cut direction. This procedure is repeated until the residuals are within the tolerance and the ellipsoid becomes sufficiently small. 

The proposed precoding design is summarized in Algorithm~\ref{alg:WMMSE_linear} and the ellipsoid method for updating Lagrange multipliers is provided in Algorithm~\ref{alg:ellipsoid}. The precoding matrices in Algorithm~\ref{alg:WMMSE_linear} are initialized using MMSE precoder as  
\begin{equation}
\label{eq:initial_precoding}
   \vect{W}_{l,k} = \sqrt{\bar{\rho}_{l,k}} \frac{\Tilde{\vect{W}}_{l,k}}{\|\tilde{\vect{W}}_{l,k}\|_F},
\end{equation}
with $\Tilde{\vect{W}}_{l,k} =  \left(\sum_{i=1}^K \Tilde{\vect{H}}_{l,i}^{\Htran} \Tilde{\vect{H}}_{l,i} + \sigma^2 \vect{I}_N\right)^{-1} \Tilde{\vect{H}}_{l,k}^{\Htran} \vect{Q}_{l,k}^{(S)}$ where $\vect{Q}_{l,k}^{(S)}$ is the matrix collecting the $S$ dominant left singular vectors of $\vect{\tilde{H}}_{l,k}$, and  $\bar{\rho}_{l,k} = \bar{\rho_l} \frac{\sqrt{\beta_{l,k}}}{\sum_{i=1}^K \sqrt{\beta_{l,i}}}$ where $\bar{\rho}_l = \min_{x}\rho_{l,x}$.

\begin{remark}
  The proposed framework can be extended to account for different hardware impairments, such as non-linear power amplifier (PA), finite resolution digital-to-analog/analog-to-digital converters, etc. 
  For example, considering PA non-linearity, the ideal transmit signal of SAT~$l$, $\sum_{i=1}^K \vect{W}_{l,i} \vect{x}_i$, is replaced by the output of the non-linear PA,  $\alpha_l \sum_{i=1}^K \vect{W}_{l,i} \vect{x}_i + \boldsymbol \xi_l$, where $\alpha_l$ is the linear gain of the PA and $\boldsymbol \xi_l$ is the non-linear distortion noise with covariance $\boldsymbol \Xi _l$. Under this model, the effective received signal in \eqref{eq:effective_received_signal} becomes
  \begin{equation}
    \tilde{\vect{y}}_k = \sum_{l = 1}^L \Tilde{\vect{H}}_{l,k} \left(\alpha_l \sum_{i=1}^K \vect{W}_{l,i} \vect{x}_i + \boldsymbol \xi_l \right) + \vect{n}_k, 
  \end{equation}
  and the MSE matrix in \eqref{eq:MSE} becomes 
  \begin{align}
    &\vect{E}_k = \vect{U}_k^{\Htran} \vect{B}_k \vect{U}_k - 2\Re \left(\vect{U}_k^{\Htran} \sum_{l=1}^L \alpha_l \Tilde{\vect{H}}_{l,k}\vect{W}_{l,k}\right) + \vect{I}_S, 
  \end{align}
  where $\vect{B}_k = \sum_{i=1}^K \sum_{l=1}^L \alpha_l^2 \tilde{\vect{H}}_{l,k} \vect{W}_{l,i} \vect{W}_{l,i}^{\Htran} \Tilde{\vect{H}}_{l,k}^{\Htran} + \sum_{l=1}^L \tilde{\vect{H}}_{l,k} \boldsymbol \Xi_l \Tilde{\vect{H}}_{l,k}^{\Htran} + \sigma^2 \vect{I}_M$. Furthermore, the power constraints in \eqref{eq:power_constraint} should be imposed on the PA output power. Therefore, \eqref{eq:power_constraint} is replaced by 
  \begin{equation}
     \mathrm{Tr}\left(\vect{A}_{l,x} \Big(\alpha_l^2 \sum_{k=1}^K \vect{W}_{l,k} \vect{W}_{l,k}^{\Htran} + \boldsymbol \Xi_l \Big) \right) \leq \rho_{l,x}. 
  \end{equation}
  Such hardware impairments generally degrade the achievable system performance by introducing additional distortion. The key point, however, is that these effects can be incorporated into the signal model, the MSE matrix, and the power constraints without changing the overall WMMSE-based structure. Hence, it is straightforward to apply the same framework for transceiver design using the modified hardware-aware expressions. 
\end{remark}

\begin{algorithm}[t]
\small
\caption{ Proposed Precoding Design under General Convex Power Constraints}
\label{alg:WMMSE_linear}
\begin{algorithmic}[1]
\STATEx{\textbf{Inputs:} $\varphi_{l,k}$, $\theta_{l,k}$, $\beta_{l,k}~\forall l,k$, $\sigma^2$, $\{\vect{A}_{l,x},\rho_{l,x}\}_{x=1}^{X_l}~\forall l$, $\epsilon$, $I_{\max}$.}

\STATE Initialize the precoding matrices $\vect{W}_{l,k}~\forall l,k$.
\STATE Initialize $\delta=\epsilon+1$ and set the initial value of the objective function in \eqref{eq:WMMSE_OF} to $+\infty$.
\STATE Set $I=0$.

\WHILE{$\delta>\epsilon ~\&~ I<I_{\max}$}
    \STATE $I\leftarrow I+1$.
    \STATE Find receive combining matrices $\vect{U}_k^\star~\forall k$ from \eqref{eq:receive_beamforming}.
    \STATE Find weight matrices $\vect{C}_k^\star~\forall k$ from \eqref{eq:weight_matrix}.
    \LongState{For each SAT $l$, obtain $\boldsymbol{\mu}_l^\star=[\mu_{l,1}^\star,\ldots,\mu_{l,X_l}^\star]^{\Ttran}$ via Algorithm~\ref{alg:ellipsoid}.}
    \LongState{Update the precoding matrices $\vect{W}_{l,k}^\star(\boldsymbol{\mu}_l^\star)~\forall l,k$ using \eqref{eq:precoding_matrix}.}
    \LongState{Compute the objective function in \eqref{eq:WMMSE_OF} with the updated variables.}
    \LongState{Set $\delta$ as the difference between the current and previous objective function values.}
\ENDWHILE

\STATEx{\textbf{Output:} $\vect{W}_{l,k}^\star(\boldsymbol{\mu}_l^\star)~\forall l,k$.}
\end{algorithmic}
\end{algorithm}

\begin{algorithm}[t]
\small
\caption{ Ellipsoid Method for Lagrange Multipliers Update (SAT~$l$)}
\label{alg:ellipsoid}
\begin{algorithmic}[1]
\STATEx{\textbf{Inputs:} $\Tilde{\vect{H}}_{l,k}$, $\vect{U}_k$, $\vect{C}_k~\forall k$, $\{\vect{A}_{l,x},\rho_{l,x}\}_{x=1}^{X_l}$, $\upsilon>1$, $\varepsilon$, $I_{\mathrm{ell}}$.}
\STATEx{\textbf{Initialization:}}
\STATE Set $d=X_l$ (dimension of $\boldsymbol{\mu}_l$) and $\boldsymbol{\mu}_l=\vect{0}$.
\STATE Compute $\vect{W}_{l,k}^\star(\boldsymbol{\mu}_l)~\forall k$ from \eqref{eq:precoding_matrix} and the residuals $g_{l,x}(\boldsymbol{\mu}_l)~\forall x$ from \eqref{eq:residuals}.
\IF{$g_{l,x}(\boldsymbol{\mu}_l)\le 0$ for all $x=1,\ldots,d$}
    \STATE \textbf{return} $\boldsymbol{\mu}_l^\star=\vect{0}$.
\ENDIF

\STATE Choose an initial vector $\boldsymbol{\mu}_l^{(0)}\succ \vect{0}$.
\WHILE{$\exists x$ such that $g_{l,x}(\boldsymbol{\mu}_l^{(0)})>0$}
    \STATE $\boldsymbol{\mu}_l^{(0)} \leftarrow \upsilon \,\boldsymbol{\mu}_l^{(0)}$  (geometric expansion)
    \LongState{ Recompute $\vect{W}_{l,k}^\star(\boldsymbol{\mu}_l^{(0)})~\forall k$ from \eqref{eq:precoding_matrix} and $g_{l,x}(\boldsymbol{\mu}_l^{(0)})~\forall x$ from \eqref{eq:residuals}.}
\ENDWHILE
\STATE Set $\bar{\boldsymbol{\mu}}_l=\boldsymbol{\mu}_l^{(0)}$ (feasible upper bound).
\STATE $\vect{c}^{(0)}=\bar{\boldsymbol{\mu}}_l/2$,~ $\vect{P}^{(0)}=\frac{d}{4}\,\mathrm{diag}(\bar{\boldsymbol{\mu}}_l\odot\bar{\boldsymbol{\mu}}_l)$.
\STATE Initialize an ellipsoid containing the box $[\vect{0},\bar{\boldsymbol{\mu}}_l]$: $\mathcal{E}^{(0)}=\{\boldsymbol{\mu}:(\boldsymbol{\mu}-\vect{c}^{(0)})^{\Ttran}(\vect{P}^{(0)})^{-1}(\boldsymbol{\mu}-\vect{c}^{(0)})\le 1\}$.

\STATEx{\textbf{Ellipsoid iterations:}}
\FOR{$t=0,1,\ldots,I_{\mathrm{ell}}-1$}
    \LongState {Set $\boldsymbol{\mu}_l^{(t)}=\vect{c}^{(t)}$ and compute $\vect{W}_{l,k}^\star(\boldsymbol{\mu}_l^{(t)})~\forall k$ from \eqref{eq:precoding_matrix}.}
    \LongState {Compute residual vector $\vect{g}_l^{(t)}=[g_{l,1}(\boldsymbol{\mu}_l^{(t)}),\ldots,g_{l,d}(\boldsymbol{\mu}_l^{(t)})]^{\Ttran}$ using \eqref{eq:residuals}.}
    \IF{$\max_x g_{l,x}(\boldsymbol{\mu}_l^{(t)})\le \varepsilon$ \& $\max_i \sqrt{\vect{P}^{(t)}(i,i)}\le \varepsilon$}
        \STATE \textbf{break}
    \ENDIF

    \STATE Set cut normal $\vect{s}^{(t)}=\vect{g}_l^{(t)}$.
    \STATE Normalize: $\tilde{\vect{s}}^{(t)}=\dfrac{\vect{P}^{(t)}\vect{s}^{(t)}}{\sqrt{(\vect{s}^{(t)})^T\vect{P}^{(t)}\vect{s}^{(t)}}}$.
    \STATE Update center (central cut):
    \STATE \hspace{0.6cm} $\vect{c}^{(t+1)}=\vect{c}^{(t)}-\dfrac{1}{d+1}\tilde{\vect{s}}^{(t)}$.
    \STATE Enforce non-negativity: $\vect{c}^{(t+1)}\leftarrow [\vect{c}^{(t+1)}]_+$.
    \STATE Update shape matrix:
    \STATE \hspace{0.6cm} $\vect{P}^{(t+1)}=\dfrac{d^2}{d^2-1}\left(\vect{P}^{(t)}-\dfrac{2}{d+1}\tilde{\vect{s}}^{(t)}(\tilde{\vect{s}}^{(t)})^T\right)$.
\ENDFOR
\STATEx{\textbf{Output:} $\boldsymbol{\mu}_l^\star=\vect{c}^{(t)}$.}
\end{algorithmic}
\end{algorithm}

\section{Fronthaul-Aware Streamwise Transmission with Limited Message Sharing}
\label{sec:streamwise}
In Section~\ref{sec:joint_precoding}, we studied a joint transmission model where each UE is served by all SATs and SAT~$l$ applies a full precoding matrix $\vect{W}_{l,k}$ to potentially transmit all $S$ streams to UE~$k$. This approach requires distributing all $S$ data streams of each UE to all SATs. In practice, such full message sharing can be challenging in LEO networks because the same data must be transferred from the ground to multiple SATs, either through dedicated links from ground stations or through inter-satellite links. In either case, this imposes additional requirements on fronthaul capacity, as well as on latency and onboard processing constraints.
Motivated by these practical limitations, we next propose a reduced-overhead scheme in which each of the $S$ streams intended for UE~$k$ is transmitted by only one SAT. In this section, we consider the per-SAT total power constraints as a special case of the general power constraints studied in Section~\ref{sec:joint_precoding}. 
\subsection{Signal Model}
Let $x_{k,s}$ denote the $s$-th data stream intended for UE~$k$ with $\mathbb{E}\{|x_{k,s}|^2\}=1$ and $\mathbb{E}\{x_{k,s}x_{i,t}^{*}\}=0$ for $(k,s)\neq(i,t)$. Let $\pi_k(s)\in\{1,\ldots,L\}$ denote the index of the SAT assigned to transmit stream~$s$ of UE~$k$, and let $\vect{w}_{\pi_k(s),k,s}\in\mathbb{C}^{N\times 1}$ denote the corresponding precoding vector.
The received signal at UE~$k$ is
\begin{align}
\label{eq:sw_received_signal_actual}
\vect{y}_k
=
\sum_{i=1}^{K}
\sum_{s=1}^{S}
\vect{H}_{\pi_i(s),k}
\vect{w}_{\pi_i(s),i,s}
x_{i,s}
+
\vect{n}_k.
\end{align}

Similar to the joint transmission case, directly optimizing the ergodic achievable SE is challenging because the channel gains contain random phases. Therefore, we adopt the same sCSI-based surrogate SE introduced in Section~\ref{sec:sysmod} and replace $\vect{H}_{l,k}$ with the deterministic effective channel $\Tilde{\vect{H}}_{l,k}$. This gives the following effective streamwise received signal:
\begin{equation}
\Tilde{\vect{y}}_k
=
\sum_{i=1}^{K}
\sum_{s=1}^{S}
\Tilde{\vect{H}}_{\pi_i(s),k}
\vect{w}_{\pi_i(s),i,s}
x_{i,s}
+
\vect{n}_k .
\label{eq:sw_received_signal_effective}
\end{equation}
The WMMSE-based streamwise precoder design is developed next based on \eqref{eq:sw_received_signal_effective}.

\subsection{Precoding Design}
\label{sec:precoding_streamwise}

With the received signal model in \eqref{eq:sw_received_signal_effective} and applying the linear combiner $\vect{U}_k$ to form $\hat{\vect{x}}_k = \vect{U}_k^{\Htran} \Tilde{\vect{y}}_k$, the MSE matrix becomes 
 \begin{align}
 \label{eq:new_MSE}
 &\vect{E}_k = \mathbb{E}\left\{ (\hat{\vect{x}}_k - \vect{x}_k)(\hat{\vect{x}}_k - \vect{x}_k)^{\Htran} \right\} = \nonumber \\
 & \vect{U}_k^{\Htran} \left (\sum_{i=1}^K \sum_{s=1}^S \Tilde{\vect{H}}_{\pi_i(s),k} \vect{w}_{\pi_i(s),i,s} \vect{w}_{\pi_i(s),i,s}^{\Htran} \Tilde{\vect{H}}_{\pi_i(s),k}^{\Htran}\right) \vect{U}_k \nonumber \\ &-2\Re\left(\vect{U}_k^{\Htran} \vect{G}_k\right) + \sigma^2 \vect{U}_k^{\Htran} \vect{U}_k + \vect{I}_S,
 \end{align}
 where the desired effective channel matrix for UE~$k$ is
 \begin{align}
   &\vect{G}_k \triangleq \nonumber \\ &\left[\tilde{\vect{H}}_{\pi_k(1),k}\vect{w}_{\pi_k(1),k,1},\ldots, \tilde{\vect{H}}_{\pi_k(S),k}\vect{w}_{\pi_k(S),k,S}\right] \in \mathbb{C}^{M \times S}.  
 \end{align}
Following the same steps as in Section~\ref{sec:joint_precoding} to cast the sum SE maximization problem as a weighted sum MSE minimization problem and applying block coordinate descent, the optimal receive combining vector and weight matrix can be obtained as 
\begin{align}
 \label{eq:receive_beamforming_streamwise}&\vect{U}_k^\star = \vect{J}_k^{-1} \vect{G}_k, \\
 \label{eq:weight_matrix_streamwise}&\vect{C}_k^\star = \frac{1}{\ln 2} \vect{E}_k^{-1},
\end{align}
where
\begin{equation}
 \vect{J}_k \triangleq \sum_{i=1}^K \sum_{s=1}^S \Tilde{\vect{H}}_{\pi_i(s),k} \vect{w}_{\pi_i(s),i,s} \vect{w}_{\pi_i(s),i,s}^{\Htran} \Tilde{\vect{H}}_{\pi_i(s),k}^{\Htran} + \sigma^2 \vect{I}_M.   
\end{equation}
Next, we need to optimize the precoding vectors having fixed $\{\vect{U}_k\}$ and $\{\vect{C}_k\}$. The corresponding optimization problem is 
\begin{subequations}
\label{eq:precoding_problem_streamwise}
 \begin{align}
  \minimize{\{\vect{w}_{l,k,s}\}_{(k,s) \in \mathcal{T}_l}}&~~\,\sum_{k=1}^K \mathrm{Tr}(\vect{C}_k \vect{E}_k), \\
    \mathrm{subject~to}~~&\sum_{(k,s) \in \mathcal{T}_l} \|\vect{w}_{l,k,s}\|^2 \leq \rho_l,~\forall l, 
    \end{align}
 \end{subequations}
 where $\mathcal{T}_l \triangleq \{(k,s): \pi_k(s) = l\}$ and $\rho_l$ denotes the maximum transmit power for SAT~$l$.
 After some mathematical manipulations, problem \eqref{eq:precoding_problem_streamwise} can be decoupled into $L$ independent sub-problems, each corresponding to precoding optimization for a particular SAT. The $l$-th sub-problem can be expressed as 
\begin{subequations}
\label{eq:precoding_subproblems}
 \begin{align}
  \minimize{\{\vect{w}_{l,k,s}\}_{(k,s) \in \mathcal{T}_l}} &\sum_{(i,s) \in \mathcal{T}_l} \left(\vect{w}_{l,k,s}^{\Htran} \vect{T}_l \vect{w}_{l,k,s} - 2\Re(\vect{z}_{l,k,s}^{\Htran} \vect{w}_{l,k,s})\right), \\
    \mathrm{subject~to}~~ & \label{eq:per_SAT_power_const}\sum_{(k,s) \in \mathcal{T}_l} \|\vect{w}_{l,k,s}\|^2 \leq \rho_l,~\forall l, 
    \end{align}
 \end{subequations}
 with 
 \begin{align}
     &\label{eq:T_l}\vect{T}_l \triangleq \sum_{i=1}^K \tilde{\vect{H}}_{l,i}^{\Htran} \vect{U}_i \vect{C}_i \vect{U}_i^{\Htran} \tilde{\vect{H}}_{l,i} \in \mathbb{C}^{N \times N}\\
     &\label{eq:z_l}\vect{z}_{l,k,s} \triangleq \tilde{\vect{H}}_{l,k}^{\Htran}\vect{U}_k \vect{C}_k \vect{e}_s \in \mathbb{C}^N,
 \end{align}
where $\vect{e}_s$ is the $s$-th standard basis vector in $\mathbb{C}^S$. Applying the Lagrange multiplier approach, the solution to \eqref{eq:precoding_subproblems} can be obtained as
\begin{equation}
\label{eq:precoding_vectors}
 \vect{w}^\star_{l,i,s}(\mu_l) = \left(\vect{T}_l + \mu_l \vect{I}_N\right)^{-1} \vect{z}_{l,i,s},   
\end{equation}
where $\mu_l$ is the Lagrange multiplier associated with constraint \eqref{eq:per_SAT_power_const} and can be updated using the bisection method. 

\subsection{Streamwise Satellite Association}
\label{sec:SW_association}
In the streamwise transmission scheme described in Section~\ref{sec:precoding_streamwise}, the $s$-th stream of UE~$k$ is transmitted by only one SAT, indexed by $\pi_k(s)$, and this mapping determines which precoding vectors are active in the received signal model \eqref{eq:sw_received_signal_effective} and in the per-SAT assignment sets $\mathcal{T}_l$. In this subsection, we specify how to construct $\{\pi_k(s)\}_{s=1}^{S}$ for UE~$k$. 

Recall that the sCSI-based channel matrix from SAT~$l$ to UE~$k$ is $\tilde{\vect{H}}_{l,k}\in\mathbb{C}^{M\times N}$. For a given UE~$k$, we form the aggregated sCSI-based channel by concatenating all $L$ blocks:
\begin{equation}
\label{eq:aggregated_channel}
\tilde{\vect{H}}_{k}\triangleq
\big[\tilde{\vect{H}}_{1,k},\tilde{\vect{H}}_{2,k},\ldots,\tilde{\vect{H}}_{L,k}\big]
\in\mathbb{C}^{M\times LN}.
\end{equation}
Since UE~$k$ has $M$ antennas, $\tilde{\vect{H}}_k$ has at most $M$ non-zero singular values (under the standard assumption that $LN \geq M$). Let the economy-size singular value decomposition (SVD) be
\begin{equation}
\tilde{\vect{H}}_{k}=\vect{Q}_{k}\vect{\Sigma}_{k}\vect{V}_{k}^{\Htran},
\label{eq:ec_svd}
\end{equation}
with $\vect{Q}_k \in \mathbb{C}^{M \times M}$, $\vect{\Sigma}_k \in \mathbb{R}^{M \times M}$, and $\vect{V}_k \in \mathbb{C}^{LN \times M}$. Denote by $\vect{v}_{k,m}$ the $m$-th column of $\vect{V}_k$. This vector describes the $m$-th transmit-side eigenmode of UE~$k$ across the aggregated antenna array formed by all SATs. To perform the streamwise association, we first quantify how the energy of $\vect{v}_{k,m}$ is distributed across the $L$ SATs' antenna arrays. In other words, we measure how much each SAT~$l$ contributes to the $m$-th eigenmode. To this end, we partition $\vect{v}_{k,m}$ into $L$ consecutive blocks of length $N$, with each block corresponding to one SAT:
\begin{equation}
\vect{v}_{k,m}=
\big[(\vect{v}^{(1)}_{k,m})^{\Ttran},\ldots,(\vect{v}^{(L)}_{k,m})^{\Ttran}\big]^{\Ttran},
~~
\vect{v}^{(l)}_{k,m}\in\mathbb{C}^{N}.
\label{eq:eigenmode_blocks}
\end{equation}
Since $\|\vect{v}_{k,m}\|^2=1$, we define the normalized contribution of SAT~$l$ to the $m$-th eigenmode as
\begin{equation}
\label{eq:participation_factor}
\eta_{l,k,m}\triangleq \big\|\vect{v}^{(l)}_{k,m}\big\|^{2},
\,\, l=1,\ldots,L,\,\, m=1,\ldots,M,
\end{equation}
which satisfies $\sum_{l=1}^{L}\eta_{l,k,m}=1$ for every $m$. Hence, $\eta_{l,k,m}$ can be interpreted as a participation factor indicating how much of the $m$-th eigenmode is carried by the antennas of SAT~$l$.

We intend to transmit $S \leq M$ streams over the $S$ strongest eigenmodes of $\Tilde{\vect{H}}_k$, and assign each of these eigenmodes/streams to a serving SAT. We next determine the stream-SAT association $\pi_k(s)$ by enforcing a one-to-one assignment between the $S$ streams of UE~$k$ and $S$ distinct SATs. To this end, we introduce the binary assignment variables $q_{l,k,s}\in\{0,1\}$, where $q_{l,k,s}=1$ indicates that SAT~$l$ is selected to transmit the $s$-th stream of UE~$k$. Using the participation factors in \eqref{eq:participation_factor}, we compute the association for UE~$k$'s streams by solving the following linear assignment problem:
\begin{subequations}
\label{eq:assignment_problem}
 \begin{align}
  \maximize{\{q_{l,k,s}\}_{\forall l,s}}~&\sum_{s=1}^{S}\sum_{l=1}^{L} \eta_{l,k,s}\, q_{l,k,s} \label{eq:assignment_obj}, \\
\mathrm{subject~to}~&\sum_{l=1}^{L} q_{l,k,s} = 1,~\forall s, \label{eq:assignment_c1}\\
& \sum_{s=1}^{S} q_{l,k,s} \le 1,~ \forall l, \label{eq:assignment_c2}\\
& q_{l,k,s}\in\{0,1\},~ \forall l,s. \label{eq:assignment_bin}
    \end{align}
 \end{subequations}
Constraint \eqref{eq:assignment_c1} ensures that each stream~$s$ is assigned to exactly one SAT, while \eqref{eq:assignment_c2} ensures that each SAT is used for at most one stream of UE~$k$, yielding $S$ distinct serving SATs (assuming $L\ge S$). Problem \eqref{eq:assignment_problem} is a maximum-weight bipartite matching problem and can be solved efficiently by applying the Hungarian algorithm \cite{Kuhn1955hungarian}. 
The resulting mapping is obtained as $\pi_k(s)=l$ if $q_{l,k,s}=1$. In a nutshell, the mapping $\pi_{k}(s)$ decides which SAT is responsible for transmitting stream~$s$ to UE~$k$, and stream~$s$ here corresponds to a specific UE spatial direction. 

Once $\{\pi_k(s)\}$ are obtained, for all $(l,k,s)$ with $l\neq \pi_k(s)$, the corresponding beamforming vector is zero. Consequently, each SAT~$l$ optimizes the subset of vectors indexed by $\mathcal{T}_l \triangleq \{(k,s):\pi_k(s)=l\}$ in the precoder optimization problem \eqref{eq:precoding_problem_streamwise}.

\begin{remark}
    There have been several efforts in terrestrial cell-free massive MIMO to reduce the fronthaul and computational burden by replacing full cooperation, where all access points serve all UEs, with user-centric clustering, where each UE is served only by a subset of access points\cite{Parida2023cell,Aboulfotouh2025optimizing}. This idea is also relevant to the system considered in this paper: instead of allowing all SATs to transmit all UE streams, as in the joint transmission strategy of Section~\ref{sec:joint_precoding}, one could assign a subset of SATs to each UE to reduce fronthaul overhead. Nevertheless, such a clustering approach would still require that every SAT in the selected cluster obtain all the UE’s data streams. In contrast, the proposed streamwise transmission strategy performs the association at the level of individual spatial streams, ensuring that each stream is shared with only one SAT. This further reduces the data-sharing overhead, although it may come at the expense of a reduced SE compared with fully joint transmission. 
    Moreover, there have been distributed precoding strategies in the cell-free massive MIMO literature that reduce the CSI-exchange burden by letting each access point design the precoder based only on locally available CSI together with limited coordination information \cite{Yi2025distributed,Zhong2022low}. These approaches mainly address the CSI-sharing bottleneck and are therefore different from the proposed streamwise scheme, which aims to address the data-sharing bottleneck.  
\end{remark}

\begin{algorithm}[!t]
\small
\caption{Streamwise SAT Association and Precoding Design }
\label{alg:streamwise}
\begin{algorithmic}[1]
\STATEx{\textbf{Inputs:} $\varphi_{l,k}$, $\theta_{l,k}$, $\beta_{l,k}~\forall l,k$, $\sigma^2$, $\rho_l~\forall l$,  $\epsilon$, $I_{\max}$.}

\STATE Form the sCSI-based effective channel matrices $\Tilde{\vect{H}}_{l,k}=\sqrt{\beta_{l,k}}\vect{b}_{l,k}\vect{a}_{l,k}^{\Ttran}\,\forall l,k$.

\FOR{$k=1,2,\ldots,K$}
    \STATE Construct the aggregated channel $\Tilde{\vect{H}}_{k}$ using \eqref{eq:aggregated_channel}.
    \STATE Compute the economy-size SVD $\Tilde{\vect{H}}_{k}=\vect{Q}_k\vect{\Sigma}_k\vect{V}_k^{\Htran}$.
    \FOR{$m=1,2,\ldots,M$}
        \STATE Partition $\vect{v}_{k,m}$  into $\{\vect{v}^{(l)}_{k,m}\}_{l=1}^L$ as in \eqref{eq:eigenmode_blocks}.
        \STATE Compute $\eta_{l,k,m}=\|\vect{v}^{(l)}_{k,m}\|^2$\, $\forall l$ as in \eqref{eq:participation_factor}.
    \ENDFOR
    \LongState{Solve the assignment problem \eqref{eq:assignment_problem} via the Hungarian algorithm to obtain $\{q_{l,k,s}\}$.}
    \STATE Set $\pi_k(s)=l$ if $q_{l,k,s}=1\,\forall s$.
\ENDFOR
\STATE Construct the per-SAT stream index sets $\mathcal{T}_l=\{(k,s):\pi_k(s)=l\}\,\forall l$. 
\STATE Initialize $\vect{w}_{l,k,s}\,\forall l,k,s$.
\STATE Initialize $\delta=\epsilon+1$ and set the initial value of the objective function to $+\infty$.
\STATE Set $I=0$.
\WHILE{$\delta>\epsilon ~\&~ I<I_{\max}$}
    \STATE $I\leftarrow I+1$.
    \STATE Find receive combining matrices $\vect{U}_k^\star~\forall k$ from \eqref{eq:receive_beamforming_streamwise}.
    \STATE Find weight matrices $\vect{C}_k^\star~\forall k$ from \eqref{eq:weight_matrix_streamwise}.
    \STATE For each SAT $l$, obtain $\mu_l^\star$ using the bisection method. 
     \LongState{Update the precoding vectors $\vect{w}_{l,i,s}^\star(\boldsymbol{\mu}_l^\star)$ $\forall l,\,(i,s) \in \mathcal{T}_l$ using \eqref{eq:precoding_vectors}.}

    \LongState{Compute the objective value with the updated variables.} 
    \LongState{Set $\delta$ as the difference between the current and previous objective values.}
\ENDWHILE

\STATEx{\textbf{Output:} Stream-SAT associations $\{\pi_k(s)\}$ and precoders $\{\vect{w}_{l,k,s}^\star\}$.}
\end{algorithmic}
\end{algorithm}

The proposed streamwise association and precoding design are summarized in Algorithm~\ref{alg:streamwise}. We initialize the precoding vectors as 
\begin{equation}
  \vect{w}_{l,k,s}^{(0)} = \sqrt{\bar{\rho}_{l,k,s}} \frac{\tilde{\vect{W}}_{l,k}\vect{e}_s}{\|\tilde{\vect{W}}_{l,k}\vect{e}_s\|},  
\end{equation}
where $\bar{\rho}_{l,k,s} = \rho_l \frac{\sqrt{\beta_{l,k}}}{\sum_{(i,s^\prime) \in \mathcal{T}_l} \sqrt{\beta_{l,i}}}$ if $(k,s) \in \mathcal{T}_l$ and $\bar{\rho}_{l,k,s} = 0$ otherwise, and $\tilde{\vect{W}}_{l,k}$ is defined after \eqref{eq:initial_precoding}. 

Algorithms~\ref{alg:WMMSE_linear} and \ref{alg:streamwise} follow a block coordinate descent structure, where the receive combiners, MSE weights, and transmit precoders are updated in an alternating manner while the other variable blocks are fixed. For a fixed set of other variables, each sub-problem admits a closed-form or optimally solvable update. Hence, the objective values of WMMSE problems are non-increasing over the iterations. Since the objective functions are bounded below, the objective values converge \cite{Tseng2001}.
\begin{figure}[t]
    \centering
    \includegraphics[width=0.9\linewidth]{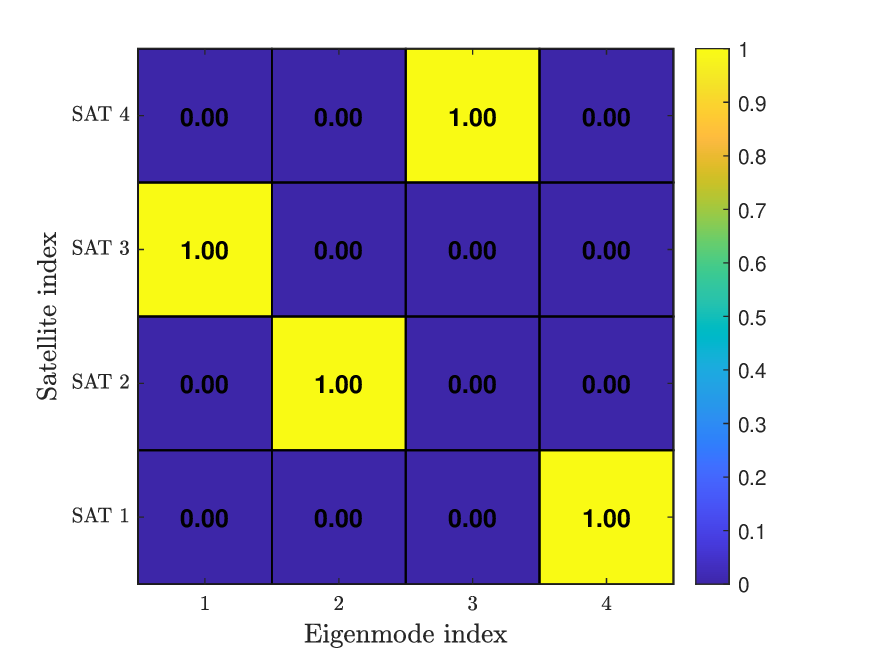}
    \caption{Satellite participation factors of the aggregated channel eigenmodes when UE-side array responses are orthogonal. In this case, each SAT contributes essentially to one eigenmode and  zero to the others. }
\label{fig:eigenmodes1}
\end{figure}

To further illustrate how the participation factors $\{\eta_{l,k,m}\}$ describe the eigenmode structure of the aggregated channel $\tilde{\mathbf H}_k$, we consider a single-UE toy example with $M=4$ antennas and $L=4$ serving SATs.  For a half-wavelength ULA with $M=4$, we have $\mathbf b(\theta)=[1, e^{j\pi\sin\theta}, e^{j2\pi\sin\theta}, e^{j3\pi\sin\theta}]^{\Ttran}$, and the inner product between two array response vectors can be written as
\begin{equation}
\mathbf b(\theta_i)^{\Htran}\mathbf b(\theta_j)
=\sum_{m=0}^{3} e^{j\pi m(\sin\theta_j-\sin\theta_i)}.
\end{equation}
Hence, two links arriving from $\theta_i$ and $\theta_j$ are orthogonal on the UE side when $\Delta\triangleq \sin\theta_j-\sin\theta_i \in \{\pm 0.5,\pm 1,\pm 1.5\}$.
Fig.~\ref{fig:eigenmodes1} shows a case where the UE-side array response vectors of the four SAT-UE links are mutually orthogonal. This is
obtained by selecting the AoAs such that $\sin\theta_{l}\in\{-0.9,-0.4,0.1,0.6\}$, for which all pairwise differences $\Delta$ belong to $\{\pm 0.5,\pm 1, \pm 1.5\}$. In this case, each SAT’s energy is concentrated in a single eigenmode, with near-zero projection onto the other modes. Consequently, streamwise transmission is naturally aligned with the channel eigen-structure and is expected to perform close to joint transmission discussed in Section~\ref{sec:joint_precoding}, while reducing fronthaul load. 

\begin{figure}
    \centering
    \includegraphics[width=0.9\linewidth]{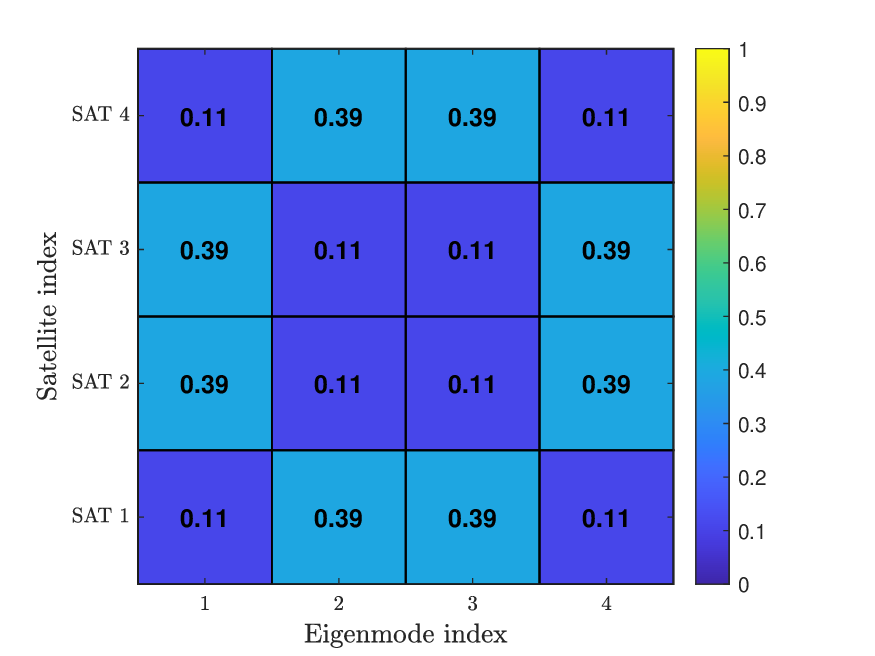}
    \caption{Satellite participation factors of the aggregated channel eigenmodes when UE-side array responses are not orthogonal. In this case, the eigenmodes are shared across multiple SATs, so each SAT contributes to more than one eigenmode. }
\label{fig:eigenmodes2}
\end{figure}

 Fig.~\ref{fig:eigenmodes2} shows a non-orthogonal case where the AoAs are set to $\theta_{l}\in\{-19.87^\circ,-6.84^\circ,6.84^\circ,19.87^\circ\}$ (equivalently, $\sin\theta_{l}\in \{-0.340,-0.119,0.119,0.340\}$), which does not satisfy the above orthogonality condition. Here, the eigenmodes are shared across multiple SATs, and the participation factors are distributed across rows, indicating that streamwise transmission departs from joint transmission and may trade SE for reduced fronthaul load. We evaluate the impact of UE-side array response orthogonality on SE in Section~\ref{sec:results}.

\begin{remark}
When the UEs are geographically close to each other, they may observe similar large-scale fading coefficients and similar dominant satellite directions. In such cases, the per-UE association problem in \eqref{eq:assignment_problem} may assign the streams of different UEs to the same subset of SATs. This can concentrate the traffic load on a few SATs while leaving other SATs unused. Moreover, if the same SAT serves closely-spaced UEs, the resulting streams may create stronger inter-user interference at the receivers. Therefore, in such scenarios, it can be beneficial to impose an additional constraint on the maximum number of streams assigned to each SAT. For example, when \(L \geq KS\), one may enforce that each SAT serves at most one stream, leading to the following global association problem:
\begin{subequations}
\begin{align}
\underset{\{q_{l,k,s}\}}{\mathrm{maximize}} \quad
& \sum_{k=1}^{K}\sum_{s=1}^{S}\sum_{l=1}^{L}
\eta_{l,k,s} q_{l,k,s} \\
\mathrm{subject~to} \quad
& \sum_{l=1}^{L} q_{l,k,s} = 1, \quad \forall k,s, \\
& \sum_{k=1}^{K}\sum_{s=1}^{S} q_{l,k,s} \leq 1, \quad \forall l, \\
& q_{l,k,s} \in \{0,1\}, \quad \forall l,k,s.
\end{align}
\end{subequations}
The resulting mapping is obtained as \(\pi_k(s)=l\) if \(q_{l,k,s}=1\). This formulation assigns each stream to a single SAT while distributing streams across distinct SATs, thereby reducing load concentration and potentially mitigating interference among closely spaced UEs.
\end{remark}

\section{Computational Complexity Analysis}
\label{sec:complexity}
In this section, we analyze the computational complexity of the proposed joint and streamwise transmission schemes.  

In the joint transmission scheme, each outer WMMSE iteration consists of three main steps: updating the receive combiners, updating the MSE weights, and updating the transmit precoders.
First, the receive combiner update in \eqref{eq:receive_beamforming} requires constructing, for each UE $k$, the $M\times M$ covariance matrix containing the contributions of all $K$ UEs and all $L$ SATs. For each $(k,i,l)$, computing  $\tilde{\mathbf{H}}_{l,k}\mathbf{W}_{l,i}$ and forming $(\tilde{\mathbf{H}}_{l,k}\mathbf{W}_{l,i})
(\tilde{\mathbf{H}}_{l,k}\mathbf{W}_{l,i})^{\Htran}$ requires $\mathcal{O}(MNS+M^2S)$ operations. Hence, forming all  covariance matrices costs $
\mathcal{O}\!\left(K^2L(MNS+M^2S)\right).$
Inverting one $M\times M$ matrix for each UE adds $\mathcal{O}(KM^3)$. The weight matrix update in \eqref{eq:weight_matrix} requires the inversion of one $S\times S$ MSE matrix per UE, which costs $\mathcal{O}(KS^3)$. 
The complexity for the transmit precoder update in \eqref{eq:precoding_matrix} is dominated by the formation of the $N\times N$ matrix $\sum_{i=1}^{K}\tilde{\mathbf{H}}_{l,i}^{\Htran}
\mathbf{U}_{i}\mathbf{C}_{i}\mathbf{U}_{i}^{\Htran}
\tilde{\mathbf{H}}_{l,i}$ and the inversion operation which,
for $L$ SATs, incur the total complexity of $\mathcal{O}\!\left(LK(NMS+NS^2+N^2S) + LN^3\right)$. Denoting by $I_{\rm ell}$ the number of iterations of the ellipsoid method, the main computational cost of one WWMSE iteration for the joint design is $
\mathcal{C}_{\rm JT}^{\rm iter}
=
\mathcal{O}\Big(
K^2LMNS + LKN^2 S
+I_{\rm ell}L N^3
\Big)$,
and the total complexity of the joint precoder design will be 
\begin{equation}
\mathcal{C}_{\rm JT}^{\rm tot} =  \mathcal{O}\left( I_{\rm J} \left(
K^2LMNS + LKN^2 S
+I_{\rm ell}L N^3 \right)
\right),
\end{equation}
where $I_{\rm J}$ is the number of WMMSE iterations for the joint transmission scheme.

The streamwise scheme has two stages: a one-time stream-SAT association stage and an iterative WMMSE precoding stage.
For the stream-SAT association, each UE first constructs the aggregated channel matrix of size $M\times LN$ and computes its economy-size SVD, costing $\mathcal{O}(M^2LN+M^3)$ per UE. The stream-SAT assignment problem is solved by the Hungarian algorithm with the complexity of $\mathcal{O}(\max\{L,S\}^3)$ per UE, which becomes $\mathcal{O}(L^3)$ when $L\geq S$. Therefore, the association stage has the dominant complexity of $
\mathcal{O}\!\left(
K\big(M^2LN+L^3\big)
\right).$

For each WMMSE iteration, the receive combiner update in \eqref{eq:receive_beamforming_streamwise} requires constructing the $M\times M$ covariance matrix at each UE. Unlike joint transmission, only one SAT transmits each stream. Hence, the covariance construction involves $KS$ active streams, and the resulting complexity is
$\mathcal{O}\!\left(K^2S(MN+M^2)\right).$
The inversion of one $M\times M$ matrix per UE adds $\mathcal{O}(KM^3)$, and the weight update costs $\mathcal{O}(KS^3)$. 
Finally, the streamwise precoder update requires constructing, for each SAT, the $N\times N$ matrix $\mathbf{T}_l$ in \eqref{eq:T_l} and the inversion of an $N \times N$ matrix, which costs
$\mathcal{O}\!\left(LK(NMS+NS^2+N^2S) + LN^3\right)$. Denoting by $I_{\rm bis}$ the required number of bisection steps, the leading-order computational complexity of one WMMSE iteration for the streamwise design is  $\mathcal{C}_{\rm SW}^{\rm iter}
=
\mathcal{O}\Big(
LKN^2S
+I_{\rm bis}L N^3
\Big)$,
and the total complexity of the streamwise precoder design is
\begin{equation}
\mathcal{C}_{\rm SW}^{\rm tot} = \mathcal{O}\left(I_{\rm S} \left(
LKN^2S
+I_{\rm bis}L N^3 \right) + K(M^2LN + L^3)
\right),
\end{equation}
with $I_{\rm S}$ being the number of WMMSE iterations for the streamwise transmission scheme. 

The streamwise scheme can incur high association complexity when the number of visible SATs $L$ is large, since the Hungarian algorithm contributes with the one-time term $\mathcal{O}(KL^3)$. However, this cost is not repeated in every WMMSE iteration. 
More importantly, the precoder and association computations are typically performed at ground gateways or network control centers with high computational capability. Therefore, in large-scale systems, the major bottleneck is often the data-sharing fronthaul rather than the computation itself. This bottleneck becomes particularly severe for joint transmission when $L$ is large, since every stream must be made available to all serving SATs, whereas streamwise transmission shares each stream with only one SAT.
The fronthaul load of the two transmission schemes will be discussed in more detail in Section~\ref{sec:results}.

  \section{Numerical Results}
\label{sec:results}
In this section, we evaluate the performance of the proposed methods for joint and streamwise transmission in distributed LEO satellite communication networks. To reflect a realistic scenario, the simulation parameters are chosen to be consistent with publicly available FCC documentation, 3GPP technical report on non-terrestrial networks (NTNs), and widely used assumptions in the literature for assessing Starlink systems \cite{FCC2021StarlinkMod,Kim2025Feasibility,3gpp-tr38821}. 

\begin{figure}
    \centering
    \includegraphics[width=0.72\linewidth]{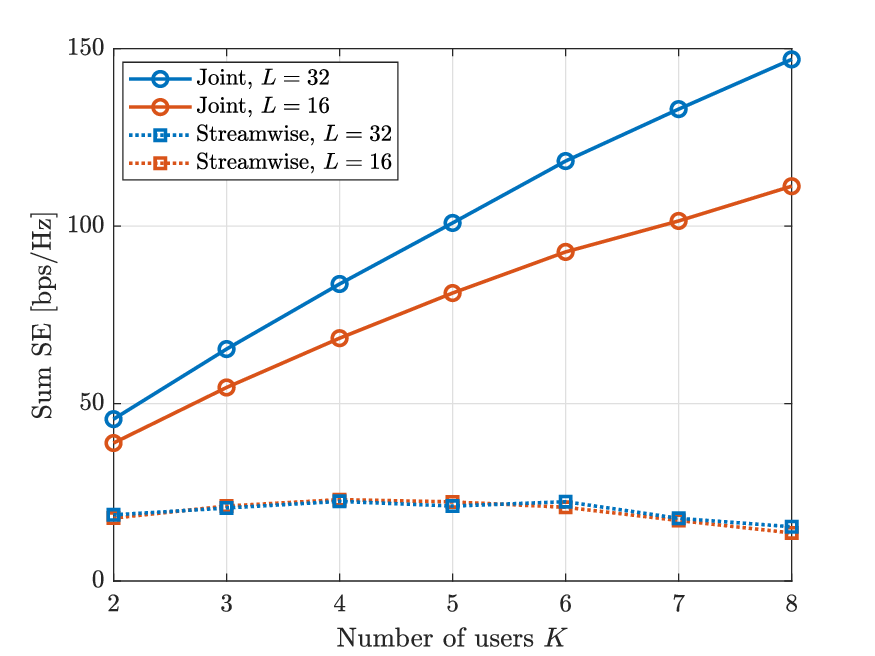}
    \caption{Performance of joint and streamwise transmissions with different numbers of UEs. } 
\label{fig:different_UEs}
\end{figure}

\begin{figure}
    \centering
    \includegraphics[width=0.72\linewidth]{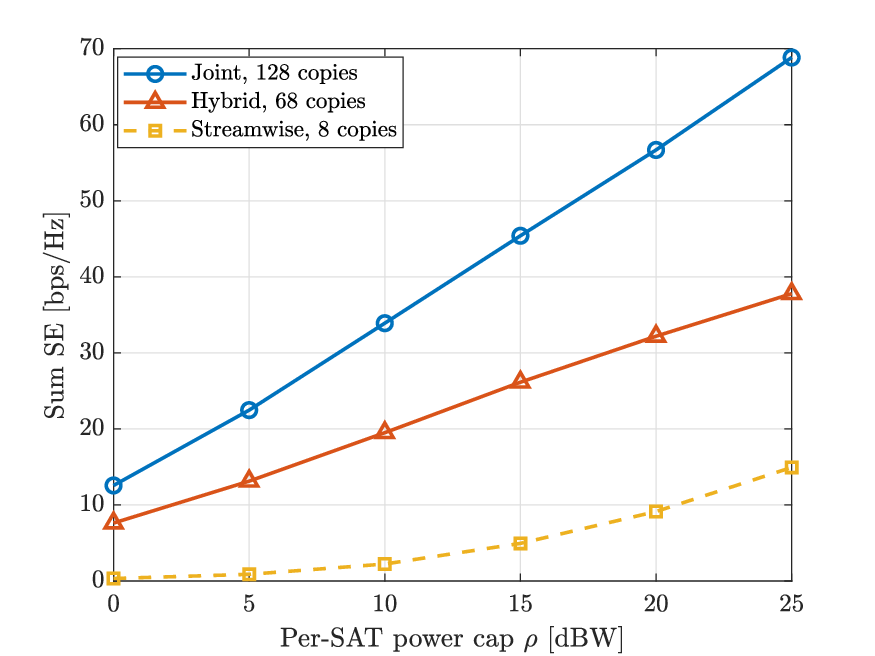}
    \caption{Comparison of  joint, streamwise, and hybrid transmission schemes.} 
\label{fig:hybrid}
\end{figure}

Unless otherwise specified, the following parameters are used for simulations: We consider an orbital altitude of $h = 560\,$km \cite{FCC2021StarlinkMod} with $L = 4$ visible SATs serving $K = 2$ UEs. Each SAT is equipped with $N = 64$ antennas and each UE has $M = 4$ antennas. Each UE is served with $S = 2$ streams.  Communication takes place at the carrier frequency of $f_c = 20\,$GHz with $400\,$MHz of bandwidth \cite{Kim2025Feasibility}, and the noise power spectral density at the UEs is $-174\,$dBm/Hz with a $1.2\,$dB noise figure \cite{3gpp-tr38821}. For the azimuth angles, we select UE~$1$ as a reference; for each SAT~$l$, the reference UE-side azimuth angles, $\theta_{l,1}~\forall l$, are independently drawn from $\mathrm{Uniform}[-90^\circ, 90^\circ]$ and the azimuth angles of UE~$2$, $\theta_{l,2}~\forall l$, are generated by adding a random drift to the reference angles, i.e., $\theta_{l,2} = \theta_{l,1} + \Delta \theta_l$, where $\Delta \theta_{l} \sim \mathrm{Uniform}[-5^\circ, 5^\circ]$ and the result is clipped to $\mathrm{Uniform}[-90^\circ, 90^\circ]$. The SAT-side azimuth angles, $\varphi_{l,k}~\forall l$, are generated similarly: for each SAT~$l$, the reference AoD $\varphi_{l,1}$ is independently drawn from $\mathrm{Uniform}[-90^\circ,90^\circ]$, and the AoD for UE~$2$ is generated as $\varphi_{l,2}=\varphi_{l,1}+\Delta\varphi_l$, where $\Delta\varphi_l\sim\mathrm{Uniform}[-5^\circ,5^\circ]$, followed by clipping to $[-90^\circ,90^\circ]$. When more than two UEs are considered, UE~1 is still used as the reference, and the angles of the other UEs are generated in the same manner. The large-scale fading coefficients are
\begin{align}
  \beta_{l,k} =  G_{\mathrm{u}} G_{\mathrm{s}} \left(\frac{\nu_c}{4 \pi f_c d_{l,k}}\right)^2, \forall l,k, 
\end{align}
where $G_{\mathrm{u}} = 8\,$dBi and $G_{\mathrm{s}} = 20\,$dBi are the antenna gains of the UEs and the SATs, respectively, and $\nu_c = 3\cdot 10^8$ is the speed of light. Furthermore, $d_{l,k}$ is the distance between SAT~$l$ and UE~$k$, given by the slant range expression as 
\begin{equation}
   d_{l,k} = - R_{\mathrm{E}}\sin(\vartheta_{l,k}) + \sqrt{(R_{\mathrm{E}}+h)^2 - (R_{\mathrm{E}}\cos(\vartheta_{l,k}))^2}, 
\end{equation}
where $R_{\mathrm{E}} = 6371\,$km is the Earth radius and $\vartheta_{l,k}$ is the elevation angle. Again, UE~$1$ is used as a reference and $\vartheta_{l,1}~\forall l$ is drawn independently from $\mathrm{Uniform}[20^\circ, 90^\circ]$.  The elevation angles for UE~$2$ are then obtained by perturbing the reference values as $\vartheta_{l,2} = \vartheta_{l,1} + \Delta \vartheta_l$ with $\Delta \vartheta_l \sim \mathrm{Uniform}[-5^\circ, 5^\circ]$, followed by clipping to the interval $[20^\circ,90^\circ]$. The Rician factor is set as $\kappa_{l,k} = 12\,$dB,\,$\forall l,k$. 
For the joint transmission scheme studied in Section~\ref{sec:joint_precoding}, we impose per-SAT power constraints. In the figures, $\rho$ denotes the common power cap applied to each SAT. In implementing the iterative algorithms, the maximum number of iterations is set to $I_{\mathrm{max}} = 40$, and the convergence threshold to $\epsilon = 10^{-4}$.
The number of Monte Carlo realizations is set to $1000$ in all simulations, except for Fig.~\ref{fig:different_UEs} and Fig.~\ref{fig:hybrid}, which involve larger system sizes. For these two simulations, the number of Monte Carlo realizations is set to $100$.

\begin{figure}
    \centering
\includegraphics[width=0.72\linewidth]{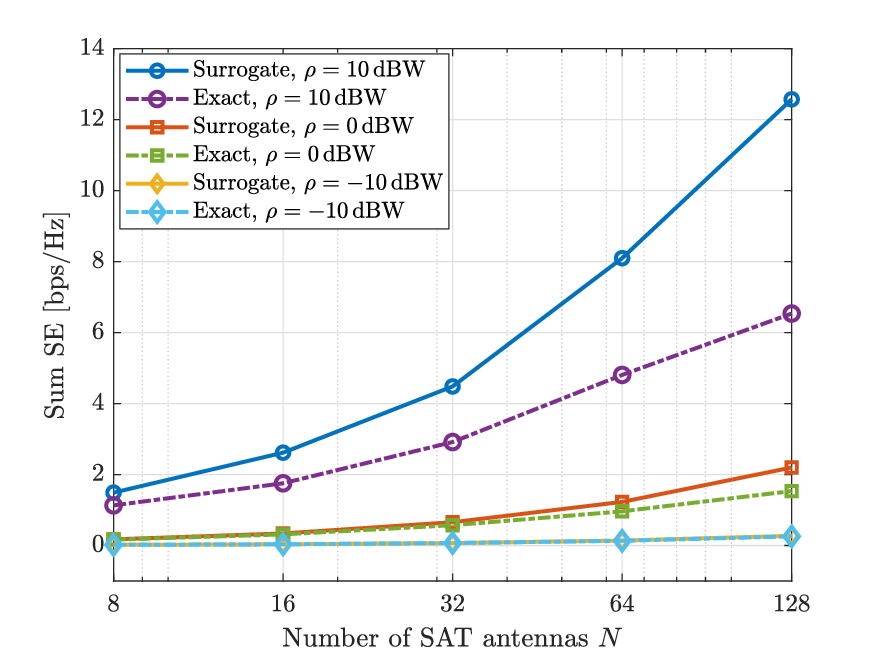}
    \caption{Comparison between surrogate and exact SE expressions.}
\label{fig:surrogate_approx}
\end{figure}

We first compare the exact ergodic SE in \eqref{eq:achievable_rate} with the surrogate SE in \eqref{eq:achievable_rate_approx}, where the precoders are obtained using Algorithm~\ref{alg:WMMSE_linear}. In Section~\ref{sec:surrogate}, we discussed two practically relevant cases in which the surrogate either acts as an upper bound on the exact expression or becomes tight with respect to it. Fig.~\ref{fig:surrogate_approx} shows the two expressions versus the number of SAT antennas $N$ for three different per-SAT power caps, namely $\rho=10$, $0$, and $-10\,$dBW. For the higher-power case, i.e., $\rho=10\,$dBW, the surrogate behaves as an upper bound on the exact SE over the considered range of $N$. This observation is consistent with the discussion in Section~\ref{sec:surrogate}. Specifically,  since the proposed precoding scheme effectively mitigates the inter-user interference, the interference-plus-noise covariance fluctuates weakly around its mean, and the surrogate expression can be interpreted as an upper bound on the exact ergodic SE. Moreover, the gap between the exact and surrogate expressions increases with $N$. This can be attributed to the additional transmit degrees of freedom provided by larger SAT arrays, which enable more effective interference suppression and therefore make the upper-bound behavior of the surrogate more pronounced. 
As the transmit power is reduced, the gap between the two expressions decreases. In particular, for $\rho=0\,$dBW, the surrogate and exact SE values become much closer, while for $\rho=-10\,$dBW, they are almost indistinguishable across the considered values of $N$. This confirms the low-power analysis in \eqref{eq:approximation_low_power}, where the desired signal power and interference become very small and the surrogate becomes a tight approximation of the exact ergodic SE.
\begin{figure}
    \centering
\includegraphics[width=0.72\linewidth]{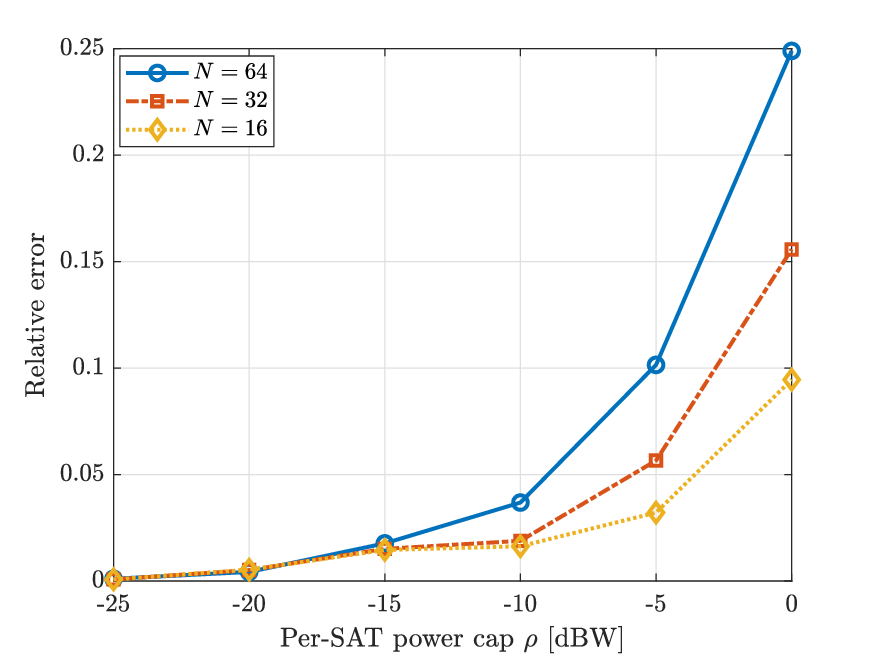}
    \caption{Relative error between the exact and surrogate SE expressions at low transmit powers.}
\label{fig:relative_error}
\end{figure}

\begin{figure}
    \centering
    \includegraphics[width=0.72\linewidth]{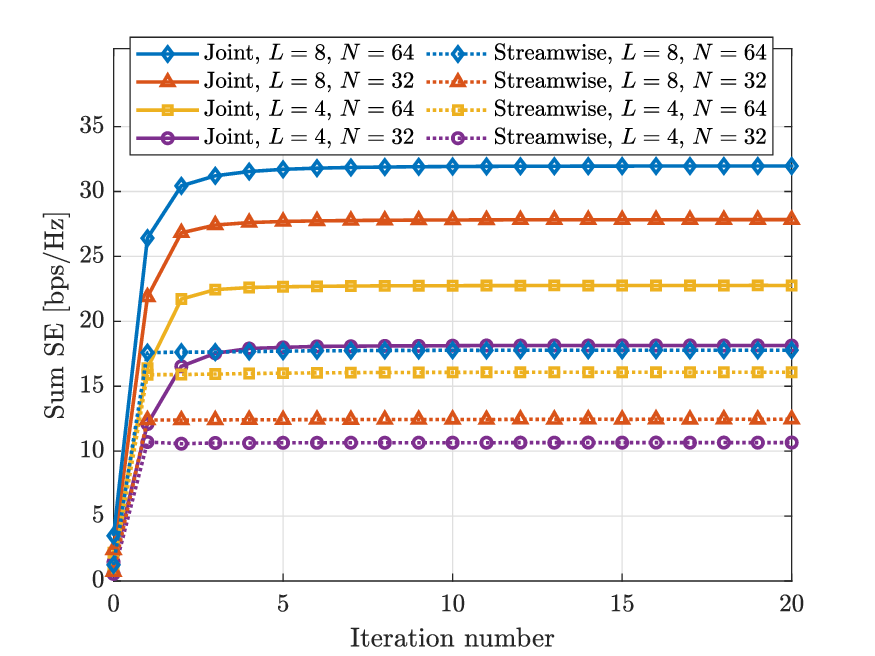}
    \caption{Convergence behavior of the proposed algorithms.}
\label{fig:convergence}
\end{figure}

\begin{figure}[t]
    \centering
    \includegraphics[width=0.72\linewidth]{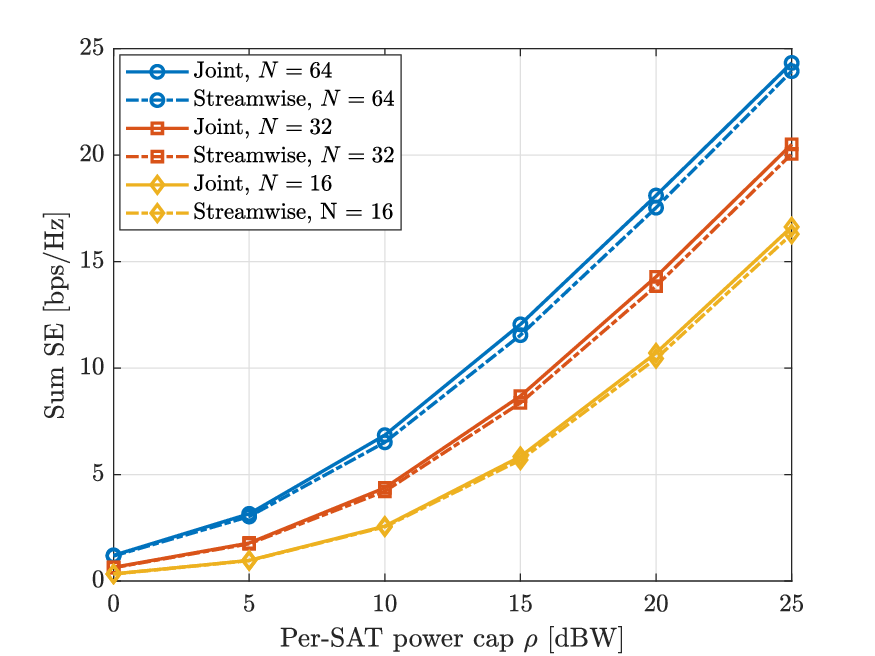}
    \caption{Comparison of joint and streamwise transmissions under UE-side orthogonality. }
\label{fig:joint_vs_sw_orth}
\end{figure}

Fig.~\ref{fig:relative_error} shows the relative error between the exact and surrogate SE expressions at low transmit powers. The relative error remains small over the considered range. Specifically, it is below $0.05$ for all considered numbers of antennas when $\rho<-10\,$dBW, and below $0.1$ when $\rho<-5\,$dBW.

Fig.~\ref{fig:convergence} demonstrates the convergence behavior of the proposed algorithms for joint and streamwise transmission when the power cap is set as $\rho = 25\,$dBW. The achieved sum SE is plotted versus the iteration number for different values of $L$ and $N$. We observe that, in all considered setups, the sum SE increases rapidly during the first few iterations and then stabilizes, indicating fast convergence of both algorithms.
\begin{figure}[t]
    \centering
    \includegraphics[width=0.72\linewidth]{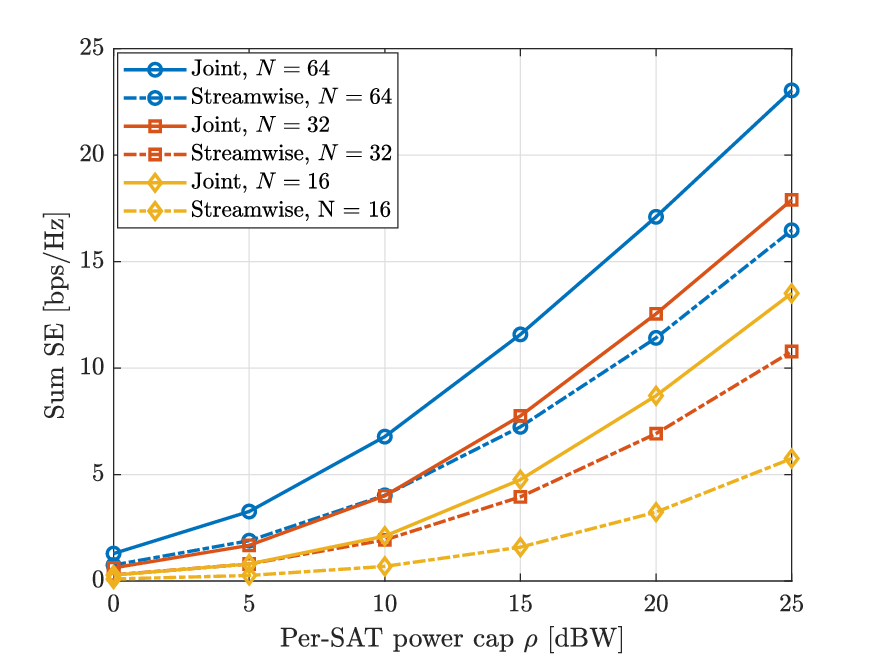}
    \caption{Comparison of joint and streamwise transmissions when UE-side channels are non-orthogonal.  }
\label{fig:joint_vs_sw_nonorth}
\end{figure}

Fig.~\ref{fig:joint_vs_sw_orth} compares the SE of the joint transmission scheme studied in Section~\ref{sec:joint_precoding} with the streamwise transmission scheme presented in Section~\ref{sec:streamwise}.
In Fig.~\ref{fig:joint_vs_sw_orth}, the UE-side azimuth angles are chosen such that, for each UE, the four SAT-UE links generate approximately orthogonal UE-side array response vectors. As a result, the aggregated sCSI-based channel $\Tilde{\vect{H}}_{k}$ exhibits an eigen-structure where each dominant UE spatial mode is carried almost entirely by one SAT. Therefore, joint transmission offers little additional gain from distributing a stream across multiple SATs, as can be observed from the figure.  In richer scattering scenarios, this gap may become slightly larger, since additional NLoS components can introduce multiple UE-side arrival directions for each SAT-UE link, so each dominant eigenmode may no longer be mainly supported by a single SAT.

Fig.~\ref{fig:joint_vs_sw_nonorth} compares joint and streamwise transmissions in the non-orthogonal setting, where the UE-side azimuth angles are randomly generated as described at the beginning of the section. In this case, the UE-side array response vectors associated with different SATs are correlated, so the SAT-UE channels are no longer separated into nearly orthogonal spatial subspaces.
Hence, joint transmission can exploit the overlap between SAT links by distributing each stream across multiple SATs and choosing the weights to shape the transmitted signals to suppress interference. On the other hand, in streamwise transmission, each stream is radiated by a single SAT, which significantly limits the ability to manage interference. As a result, streamwise transmission becomes noticeably more interference-limited, while joint transmission retains additional degrees of freedom to balance signal enhancement and interference mitigation, leading to the large SE gap between the two schemes as observed in Fig.~\ref{fig:joint_vs_sw_nonorth}. Furthermore, increasing the number of SAT antennas improves the SE of both joint and streamwise transmission since both schemes benefit from the additional SAT-side array gain. However, the gap between them is mainly governed by the UE-side spatial structure. Since increasing $N$ changes the SAT-side array response but not the UE-side angular correlation among the SAT-UE links, it mainly provides a logarithmic SE shift to both schemes. Therefore, the absolute gap between the two schemes remains nearly unchanged, although the relative loss of streamwise transmission decreases as $N$ increases.

Comparing Fig.~\ref{fig:joint_vs_sw_orth} and Fig.~\ref{fig:joint_vs_sw_nonorth}, we conclude that when the UE-side channel subspaces are close to orthogonal, streamwise transmission is near-optimal and offers a low-overhead alternative to joint transmission. When the channels are not orthogonal, joint transmission provides significant SE gains, at the cost of higher fronthaul overhead. Specifically, for the considered setup with $K = 2$, $S = 2$, and $L = 4$, streamwise transmission requires only $KS = 4$ stream copies to be shared with the SATs, while joint transmission requires $LKS = 16$ stream copies, corresponding to a $300\%$ increase in fronthaul load. 
Equivalently, if the $s$-th stream of UE~$k$ carries $r_{k,s}\,$bit/s, the data sharing fronthaul load is $\sum_{k,s}r_{k,s}$ for streamwise transmission and $L\sum_{k,s}r_{k,s}$ for joint transmission. Thus, for $L = 4$, joint transmission requires four times the data sharing bit rate of streamwise transmission. For example, if each stream carries $1\,$bps/Hz over the considered $400\,$MHz bandwidth, then $r_{k,s} = 400\,$Mbit/s. With $K = 2$, $S = 2$, and $L = 4$, streamwise transmission requires $KSr_{k,s} = 1.6\,$Gbit/s, whereas joint transmission requires $LKSr_{k,s} = 6.4\,$Gbit/s,  corresponding to an additional $4.8\,$Gbit/s data-sharing load. 
This data-sharing overhead gap grows linearly with $K$, $S$, and $L$, highlighting that the transmission mode should be selected based on a careful performance-fronthaul trade-off analysis for the target deployment. Another important benefit of streamwise transmission is that, by avoiding the additional data-sharing load, it can also reduce the fronthaul latency when the feeder link or inter-satellite link used for stream sharing becomes a bottleneck, since fewer stream copies need to be delivered before transmission.

\begin{figure}
    \centering
    \includegraphics[width=0.72\linewidth]{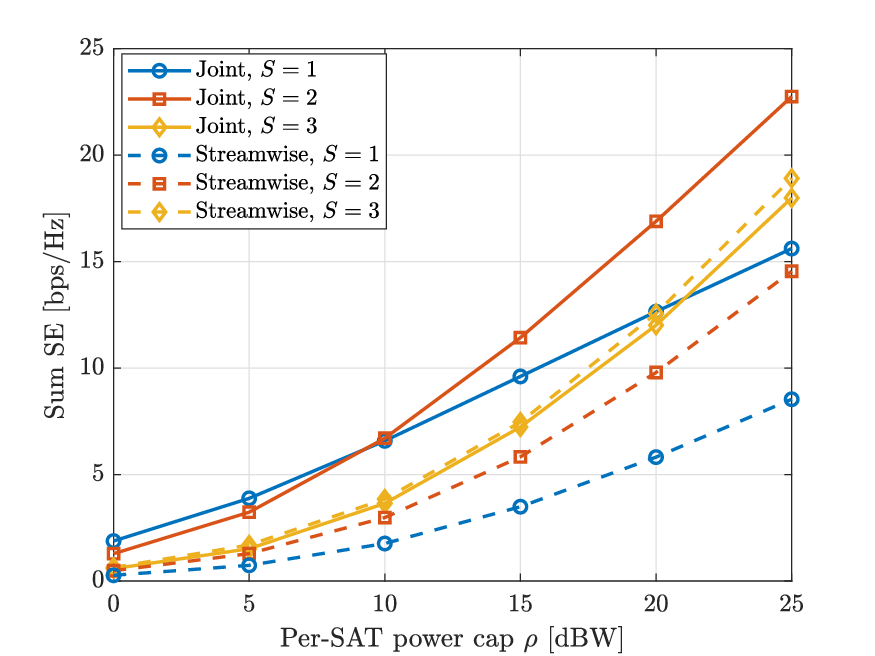}
    \caption{Performance of joint and streamwise transmissions with different numbers of streams per UE.  }
\label{fig:different_streams}
\end{figure} 

Fig.~\ref{fig:different_streams} depicts the performance of the proposed joint and streamwise transmission modes for different numbers of streams per UE. For joint transmission, increasing the number of streams from $S = 1$ to $S = 2$ improves the sum SE, particularly at higher power budgets.  However, further increasing to $S = 3$ degrades  performance since inter-user interference becomes dominant and a UE with $M = 4$ antennas has limited spatial degrees of freedom to suppress it.  
On the other hand, for streamwise transmission, increasing the number of streams from $S = 1$ to $S = 2$ and from $S = 2$ to $S = 3$ consistently improves performance as restricting each stream to a single SAT reduces the amount of interference compared to joint transmission. 
Another important observation is that, for the considered setup, joint and streamwise transmission achieve almost identical performance when $S = 3$ because the potential beamforming gain of joint transmission is largely offset by the resulting interference. 
Overall, these results highlight that the number of streams and the transmission mode should be selected jointly based on the system configuration.

Fig.~\ref{fig:different_UEs} compares joint and streamwise transmissions as the number of UEs varies for $L = 16$ and $L = 32$. The per-SAT power cap is fixed to $\rho=25\,$dBW. The joint scheme benefits from adding more UEs, since it can exploit all SATs for interference shaping and spatial multiplexing, leading to a monotonic increase in the sum SE. In contrast, streamwise transmission has limited interference mitigation capability because each stream is sent by only one SAT; hence, its SE initially increases with $K$, then saturates and eventually decreases as inter-user interference becomes severe. Increasing $L$ from $16$ to $32$ is not very helpful for streamwise transmission, since it mainly improves selection diversity and the selected SATs are already strong when $L = 16$. Joint transmission, however, uses all available SATs simultaneously, so the larger distributed aperture and additional spatial degrees of freedom stemming from having more SATs improve its performance.

To balance the trade-off between fronthaul overhead and performance when the network size grows, it is useful to consider hybrid transmission schemes that avoid the substantial fronthaul load of joint transmission, while providing better interference suppression capability than streamwise transmission. 
In Fig.~\ref{fig:hybrid}, we evaluate one such hybrid scheme when $K = 4$ and $L = 16$. In this scheme, at each Monte Carlo realization, two randomly selected UEs are served via joint transmission from all $16$ SATs, while the remaining two UEs are served using streamwise transmission. Therefore, the reported SE of the hybrid scheme is averaged over both the random channel/geometry realizations and the random selection of the UEs served by joint transmission.
With $S=2$ streams per UE, joint transmission requires $LKS=128$ stream copies, streamwise transmission requires only $KS = 8$ copies, and the considered hybrid scheme requires $KS(L+1)/2 = 68$ copies. 
We can see that the hybrid scheme outperforms streamwise transmission, while still achieving lower SE than full joint transmission. Hence, the hybrid scheme provides a balanced trade-off between fronthaul load and performance.
This motivates further investigation into the design of hybrid transmission strategies that can adaptively select which UEs or streams should be served jointly and which should use streamwise transmission.
\begin{figure}
\centering
\subfloat[Performance of the proposed scheme, MMSE precoding, and ZF precoding.]
{\includegraphics[width=0.72\linewidth]{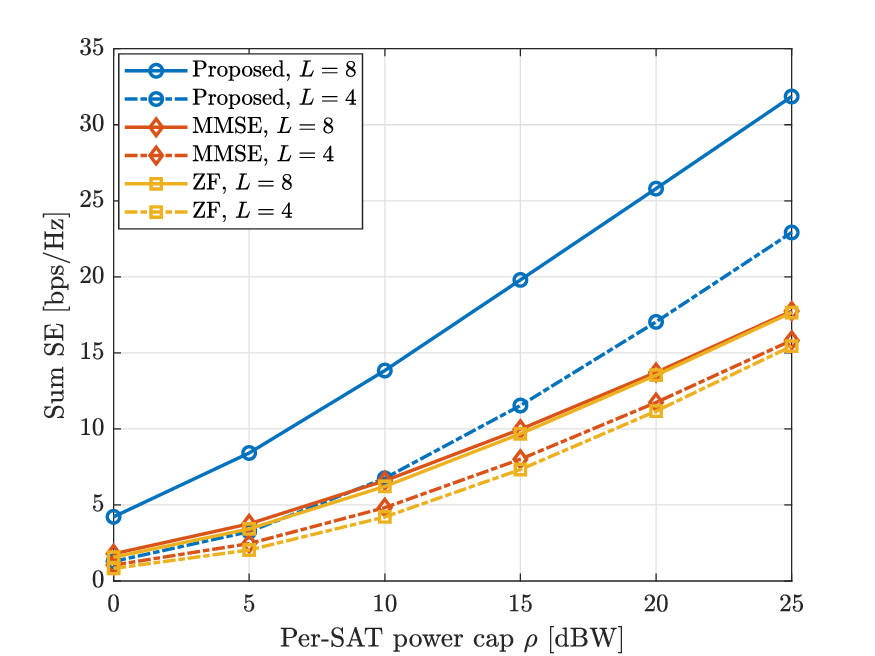}}\hfill
\centering
\subfloat[Performance of the proposed scheme and an MRT baseline with orthogonal (e.g., TDMA) transmissions, where each UE is served by its closest SAT.]
{\includegraphics[width=0.72\linewidth]{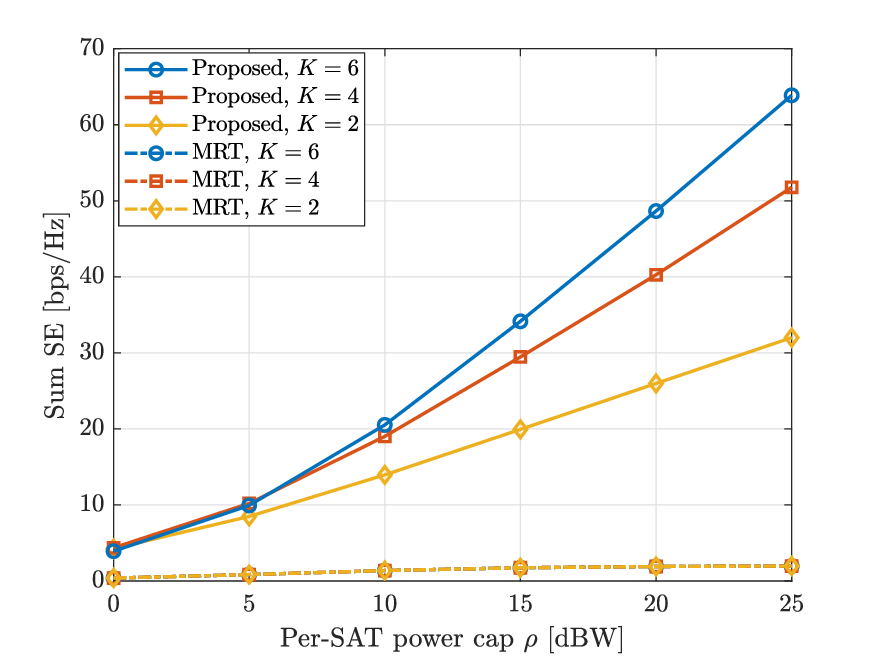}}
\caption{Performance comparison between the proposed joint precoding and well-known benchmarks.}
\label{fig:prop_vs_bench}
\vspace{-5mm}
\end{figure}

Fig.~\ref{fig:prop_vs_bench} reports the achievable SE for the proposed joint precoding in Section~\ref{sec:joint_precoding} and classical baselines. Existing cooperative satellite precoding methods in \cite{Richter2020Downlink,humadi2024distributed,ha2024user, wang2025multiple, Zhang2025Enabling} target single-antenna ground UEs and thus, do not address the multi-antenna UE setting considered in this paper and their proposed approaches are not directly comparable with the presented joint precoding design. For this reason, we benchmark against the well-known MMSE, zero forcing (ZF), and maximum ratio transmission (MRT) precoders, implemented using the available sCSI-based channel model and then enforced to satisfy per-SAT power caps. We can see in Fig.~\ref{fig:prop_vs_bench}(a) that the proposed joint transmission scheme notably outperforms MMSE and ZF benchmarks. That is because our method alternates between updating the transmit precoders, the receive combiners, and the MSE weights.
So, each iteration refines the beamformers toward directions that better suppress interference while keeping strong alignment with the desired channels. Furthermore, increasing the number of SATs from $L = 4$ to $L = 8$ provides additional distributed transmit dimensions that our iterative joint design can actively exploit to reshape interference and redistribute each SAT's power across UEs/streams and spatial directions. 
Fig.~\ref{fig:prop_vs_bench}(b) compares the  proposed joint transmission scheme with an orthogonal MRT baseline where UEs are served in an orthogonal manner by only their closest SAT, which applies MRT precoding.  We have $L = 8$ SATs and plot the results for $K = 2,\,4,\,6$ UEs. The MRT performance is poor because each UE is served by only a single SAT, and the orthogonal scheduling incurs a $1/K$ penalty.  
For the proposed joint transmission, the sum SE increases as the number of UEs grows from $K = 2$ to $K = 4$ and then to $K = 6$, since serving more UEs enables higher multi-user spatial multiplexing gains. However, the improvement from $K=4$ to $K=6$ is smaller than that from $K=2$ to $K=4$, because inter-user interference becomes more pronounced as more UEs are served simultaneously.
\begin{figure}
    \centering
    \includegraphics[width=0.72\linewidth]{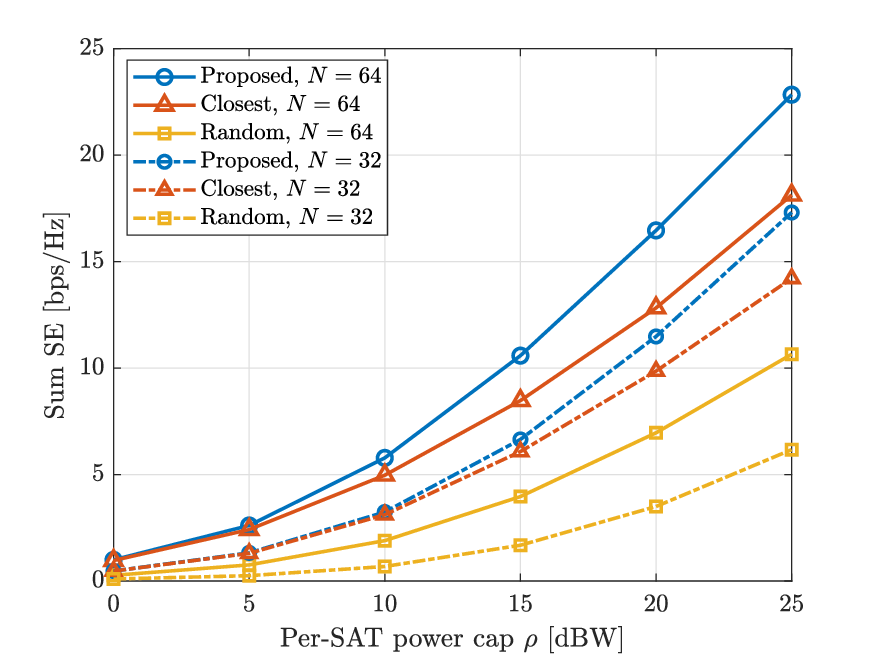}
    \caption{Performance of streamwise transmission with different stream-SAT association methods.}
\label{fig:prop_vs_random_assoc}
\end{figure}

Fig.~\ref{fig:prop_vs_random_assoc} 
evaluates the impact of stream-SAT association in the streamwise transmission studied in Section~\ref{sec:streamwise} by comparing the proposed association strategy with two benchmark schemes: a closest-SAT baseline, where the streams of each UE are assigned to the SATs with the largest large-scale fading coefficients, and a random baseline, where SATs are randomly assigned to UE eigenmodes. Specifically, in the closest-SAT baseline, for each UE, the SATs are sorted according to their large-scale fading coefficients. Then, the $S$ streams of each UE are assigned to the $S$ SATs with the largest large-scale fading coefficients.
For this simulation, $L = 8$ and the angular drifts used to generate UE~2 are increased to $10^\circ$, i.e., $\Delta \theta_l, \Delta \varphi_l, \Delta \vartheta l \sim \mathrm{Uniform}[-10^\circ, 10^\circ]$, to create a more distinguishable association scenario and avoid cases where the same SAT is closest to most UEs.
The proposed association outperforms both benchmarks because it assigns each stream based on the SAT participation factors of the corresponding channel eigenmode. In contrast, the closest-SAT baseline only accounts for link strength and ignores the eigenmode structure, while the random baseline does not exploit either channel strength or eigenmode information.
The gap between the proposed scheme and the benchmarks increases as the per-SAT power cap increases, since the system transitions from a noise-limited regime to an interference-limited regime. In this regime, the stream-SAT assignment plays a more important role in managing interference, and the proposed association becomes more beneficial. Furthermore, as the number of SAT antennas increases, the available array gain becomes larger and the proposed association can better exploit it.

\begin{figure}
    \centering
    \includegraphics[width=0.72\linewidth]{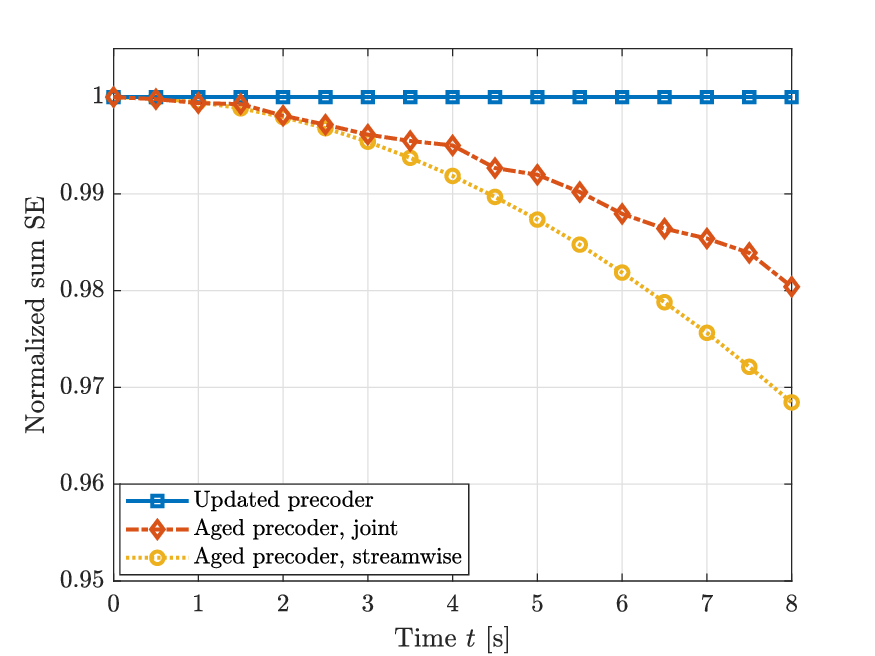}
    \caption{SE performance drop caused by aged precoder.}
\label{fig:aging}
\end{figure}

Fig.~\ref{fig:aging} illustrates the impact of precoder aging caused by SATs' orbital motion. The per-SAT power cap is fixed to $\rho=25\,$dBW. Since the proposed precoding design is based on sCSI, the relevant aging effect is the mismatch between the sCSI used for precoder computation and the updated sCSI after SAT movement. To model this effect, the AoA $\theta_{l,k}$, AoD $\varphi_{l,k}$, and elevation angle $\vartheta_{l,k}$ are updated as $\theta_{l,k}(t)=\theta_{l,k}(0)+s_{\theta,l}\eta_{\theta}\omega_s t$, $\varphi_{l,k}(t)=\varphi_{l,k}(0)+s_{\varphi,l}\eta_{\varphi}\omega_s t$, and $\vartheta_{l,k}(t)=\vartheta_{l,k}(0)+s_{\vartheta,l}\eta_{\vartheta}\omega_s t$, where $s_{\theta,l},s_{\varphi,l},s_{\vartheta,l}\in \{-1,+1\}$ denote the SAT-dependent drift directions, $\eta_{\theta}=0.5$, $\eta_{\varphi}=0.5$, and $\eta_{\vartheta}=0.3$ are drift factors, and $\omega_s=v_s/(R_{\mathrm{E}} + h)$ is the angular speed with $v_s=7.8$~km/s as the orbital speed. 
We consider an interval of $8$ seconds with a step size of $0.5$ seconds, and compare the aged precoder, computed once at the beginning and reused afterward, with the updated precoder, recomputed using the current sCSI at each time instant. The plotted metric is normalized by the SE obtained with the updated precoder for both joint and streamwise transmission. 
Hence, the updated precoder curve is equal to one, while the aged precoder curves quantify the relative SE loss due to outdated sCSI. Using $0.99$ as the performance threshold, the joint and streamwise cases drop below this threshold after approximately $5.5$ and $4.5$ seconds, respectively. Since a standard 5G NTN radio frame has a duration of $10$ ms, these update intervals correspond to about $550$ and $450$ radio frames, respectively. This shows that the proposed schemes are robust against sCSI aging and do not require frequent precoder recomputation.

\begin{figure}
    \centering
    \includegraphics[width=0.72\linewidth]{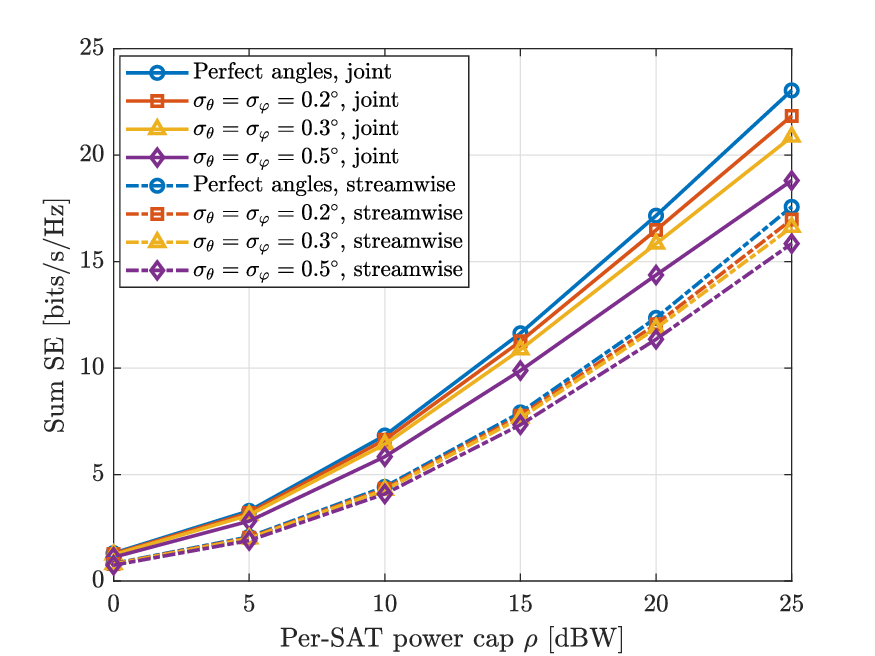}
    \caption{Performance in the presence of angle estimation errors.}
\label{fig:angle_errors}
\end{figure}

Finally, to evaluate the robustness of the proposed schemes against imperfect sCSI, Fig.~\ref{fig:angle_errors} considers angular estimation errors in the AoA and AoD used for precoder design. Specifically, the estimated AoA and AoD are obtained by perturbing their true values with independent zero-mean Gaussian errors, where $\sigma_{\theta}=\sigma_{\varphi}\in \{0.2^\circ,0.3^\circ,0.5^\circ\}$, while the resulting SE is evaluated using the true channel.
Several observations can be made from this simulation. First, streamwise transmission is more robust against angular estimation errors because each stream is transmitted by only one SAT; hence, its precoding vector is mainly affected by the angular mismatch of the assigned SAT-UE link rather than the errors from all SATs. In contrast, joint transmission relies on the angular information of all SAT-UE links, so angular errors across multiple SATs can jointly affect the desired signal and interference suppression. Second, the joint transmission scheme with angle errors of $0.2^\circ$ and $0.3^\circ$, as well as the streamwise transmission scheme for all considered error levels, exhibit only small performance degradation compared with the perfect angles case. This shows that the proposed schemes are reasonably robust to moderate sCSI estimation errors. We would also like to emphasize that in LEO satellite systems, the angular information is mainly geometry-dependent and can be accurately inferred from ephemeris data and UE location information.

\section{Conclusions}
\label{sec:conclusions}
This paper studied cell-free massive MIMO-inspired downlink transmission from multiple LEO satellites to multi-antenna ground users. Due to the limited availability of instantaneous CSI and the intractability of the exact ergodic SE, we adopted a tractable surrogate SE expression that enables precoder design using only sCSI. We considered two multi-stream transmission strategies: joint non-coherent transmission, where each satellite transmits all user streams, and streamwise transmission, where each stream is transmitted by a single satellite. None of the strategies requires phase synchronization across the satellites, in contrast to ground-based cell-free massive MIMO.

Numerical results demonstrated that streamwise transmission is nearly lossless when the UE-side SAT channels are close to orthogonal, while joint transmission provides clear gains in non-orthogonal and interference-limited regimes at the cost of higher fronthaul load. Moreover, the simulations highlighted the importance of properly selecting the number of streams and served users, and demonstrated the performance-fronthaul trade-off of a hybrid transmission strategy where only a subset of UEs is served through joint transmission while the remaining UEs use streamwise transmission to reduce the data-sharing load. Our evaluations also showed that the proposed sCSI-based designs are robust to precoder aging and moderate angular estimation errors.

\bibliographystyle{IEEEtran}
\bibliography{refs} 
\end{document}